\documentclass[11pt,a4paper]{article} 
\usepackage{graphicx}
\usepackage{jcappub}

\newcommand{\pd}{\partial}  
\newcounter{ichi}
\newcounter{ni}
\newcounter{san}
\newcounter{yon}
\makeatletter
\newsavebox{\@parc@ption}
\def\parcaption#1{%
\sbox{\@parc@ption}{\shortstack[l]{#1}}%
>\setbox\@tempboxa\hbox{\csname fnum@\@captype\endcsname}%
\@tempdima\columnwidth \advance\@tempdima-\wd\@tempboxa
\@tempdimb\@tempdima 
\ifdim\wd\@parc@ption>\@tempdimb \@tempdima\@tempdimb
\else\@tempdima\wd\@parc@ption\fi
\sbox{\@tempboxa}{\parbox[t]{\@tempdima}{#1}}%
\caption{\usebox{\@tempboxa}}}
\makeatother

\title{Galaxy clusters as reservoirs of heavy dark matter and high-energy cosmic rays: constraints from neutrino observations}

\author{Kohta Murase$^{1,2}$ and John F. Beacom$^{1,2,3}$}

\affiliation{
$^{1}$CCAPP, OSU, 191 W. Woodruff Ave., Columbus, Ohio 43210, USA\\
$^{2}$Department of Physics, OSU, 191 W. Woodruff Ave., Columbus, Ohio 43210, USA\\
$^{3}$Department of Astronomy, OSU, 140 W. 18th Ave., Columbus, Ohio 43210, USA
}

\emailAdd{murase@ias.edu, beacom.7@osu.edu}

\abstract{
Galaxy Clusters (GCs) are the largest reservoirs of both dark matter and cosmic rays (CRs).  Dark matter self-annihilation can lead to a high luminosity in gamma rays and neutrinos, enhanced by a strong degree of clustering in dark matter substructures.  Hadronic CR interactions can also lead to a high luminosity in gamma rays and neutrinos, enhanced by the confinement of CRs from cluster accretion/merger shocks and active galactic nuclei.
We show that IceCube/KM3Net observations of high-energy neutrinos can probe the nature of GCs and the separate dark matter and CR emission processes, taking into account how the results depend on the still-substantial uncertainties.  Neutrino observations are relevant at high energies, especially at $\gtrsim 10$~TeV.  Our results should be useful for improving experimental searches for high-energy neutrino emission.  Neutrino telescopes are sensitive to extended sources formed by dark matter substructures and CRs distributed over large scales.
Recent observations by \textit{Fermi} and imaging atmospheric Cherenkov telescopes have placed interesting constraints on the gamma-ray emission from GCs.  We also provide calculations of the gamma-ray fluxes, taking into account electromagnetic cascades inside GCs, which can be important for injections at sufficiently high energies.  This also allows us to extend previous gamma-ray constraints to very high dark matter masses and significant CR injections at very high energies.
Using both neutrinos and gamma rays, which can lead to comparable constraints, will allow more complete understandings of GCs.  Neutrinos are essential for dark matter annihilation channels like $\chi \chi \rightarrow \mu^+ \mu^-$, where the neutrino signals are larger than the gamma-ray signals, and for hadronic instead of electronic CRs, because only the first leads to neutrinos.  Our results suggest that the multi-messenger observations of GCs will be able to give useful constraints on specific models of dark matter and CRs.
}


\begin{document}
\maketitle

\flushbottom

\section{Introduction}
Galaxy clusters (GCs) are the largest virialized objects in the universe~\citep[see reviews][]{voi05,dia+08}.  Their mass ($\sim {10}^{14}-{10}^{15}~M_{\odot}$) consists mainly of dark matter.  Many galaxies (typically $\sim 50-1000$), including a couple of active galactic nuclei (AGN), are contained in GCs~\cite[e.g.,][]{mar+07}.  
GCs are also sources of X-rays that are mainly produced via thermal bremsstrahlung emission from the hot intracluster medium (ICM).  

The existence of dark matter is known by its gravitational effects~\cite{zwi33}, but its particle properties are unknown.  Many dark matter models are motivated by elementary particle physics, and the most popular candidate is weakly interacting massive particles (WIMPs)~\cite{dm}.  In supersymmetric models~\cite{susy}, the lightest supersymmetric particle can be stable, which is achieved by a discrete symmetry such as R-parity, although meta-stable dark matter is also allowed as long as the lifetime is sufficiently longer than the age of the universe.  
The theoretical range of WIMP annihilation cross sections (which extends over many orders of magnitude) covers values required for the correct thermal relic density.
An important way to look for dark matter signatures is by observing indirect gamma-ray and neutrinos signals from self-annihilation.  These may be seen from various astrophysical objects, including the Galactic center, nearby galaxies, dwarf galaxies, and isotropic diffuse backgrounds~\citep[e.g.,][and references therein]{dm2}.  The annihilation signal is proportional to the density squared, and not only the smooth dark matter halo but also substructures are relevant~\cite{annsubearly,annsub,annsub2,annsub3}.  

GCs can also be attractive sources to look for annihilation signatures of dark matter~\citep[e.g.,][]{col+06,pro08,jel+09}.   GCs are the largest gravitationally-bound objects in the universe, and are still forming today.  Recent N-body simulations have indicated that small, bound structures in larger halos are less affected by tidal stripping than in smaller halos, and a large population of dark matter substructures is present~\cite{annsub2,annsub3}.  This may be the case especially at the cluster outskirts~\cite{gao+11}.  If the smallest halo mass is order of the Earth mass, as suggested by arguments of the kinetic decoupling of WIMPs in the early universe~\citep[e.g.,][]{minimum}, then the annihilation signal can be boosted by a factor of $\sim 1000$ for massive GCs though more conservative values are also allowed~\cite{gao+11,pin+09,san+11,pin+11,hua+12,han+12,an12}.  This strongly motivates searches for annihilation signals from GCs.  Recently, \textit{Fermi} has started to give stringent constraints on the annihilation cross section for low-mass WIMPs~\cite{hua+12,han+12,an12,ack+10,yua+10,col+11,nez+12}.  As indicated in Table~1, one of the key points in gamma-ray and neutrino searches from GCs is the large boost factor allowed by the existence of substructures.  

GCs are also the largest cosmic-ray (CR) reservoirs in the universe. Various sources have been considered in the literature, including AGN~\cite{ber+97,fuj+07,mur+08}, supernova remnants~\cite{vol+97} and ram-pressure striping of infalling galaxies.  
Among many possibilities, AGN jets and shocks associated with the large scale structure formation seem the most promising~\cite{nor+95,cb98,merger,kw09}.  In hierarchical scenarios of cosmological structure formation, GCs should continue to grow through merging and accretion of dark matter and baryonic gas, thereby generating powerful shock waves on $\sim 1$~Mpc scales~\cite{byk+08,lw00,kes10,ryusim,min03,pfrsim,pp10,ski+08,vaz+08}.  At GC outskirts, around the virial radius, filaments and voids, shocks with relatively high Mach numbers are expected due to accretion shocks and minor merger events, and efficient acceleration of high-energy particles is possible~\cite{lw00,ryusim,ski+08,vaz+08}.  Low Mach number shocks, involving merger and flow shocks, occur inside the virial radius, which is important for thermalization~\cite{merger,pfrsim,ski+08,vaz+08}. 

\begin{table}[bt]
\begin{center}
\caption{\label{Tab1} 
Comparison among dwarf galaxies, normal galaxies and GCs.  For dark matter annihilation, GCs have an advantage in the boost factor since many substructures may survive.  The boost factor is estimated based on Ref.~\cite{gao+11}, but one should keep in mind that it has large uncertainty.  
For CRs, compared to galaxies, GCs have an advantage that higher-energy CRs can be confined because of their larger size.  Note that the coherent length is smaller than $r$, and the confinement energy is lower than $e B r$.  The magnetic field is assumed to be $1~\mu {\rm G}$ for simplicity.  Although the numbers are crude, this table demonstrates the importance of GCs for probing high-energy emissions.}
\begin{tabular}{|c||c|c|c|c|}
\hline Source & mass [$M_{\odot}$]  & Boost factor & size [pc]   & $E=e B r$ [eV]\\
\hline
\hline Dwarf Galaxies & $\sim {10}^{7}$ & $\sim 1$ & $\sim 500$ & $\sim 5 \times {10}^{17}$\\
\hline Normal Galaxies & $\sim {10}^{12}$ & $\sim 80$ & $\sim 2 \times {10}^{4}$ & $\sim 2 \times {10}^{19}$\\
\hline Clusters of Galaxies & $\sim {10}^{15}$ & $\sim 1000$ & $\sim {10}^{6}$ & $\sim {10}^{21}$\\
\hline
\end{tabular}
\end{center}
\end{table}

These CRs interact with the ICM via hadronuclear processes such as the $pp$ reaction, leading to gamma rays and neutrinos~\cite{ber+97,cb98}.  CRs may be accelerated above $\sim {10}^{17}$~eV~\cite{mur+08} and even up to ultra-high energies (UHE; $>{10}^{18.5}$~eV)~\cite{uhecr}, and they might contribute to the observed CR flux at the Earth.  CRs with $\sim {10}^{18}-{10}^{19}$~eV interact with cosmic microwave background (CMB) photons via the Bethe-Heitler process, which leads to GeV-TeV gamma rays via the electromagnetic cascade in GCs~\cite{ino+05,kot+09}.  Photomeson production, which leads to gamma-ray and neutrino production, may also happen if even higher-energy CRs exist in GCs~\cite{dem+05,kot+09,wol+08}.  Predictions for the associated nonthermal gamma-ray radiation, including leptonic emissions, have been discussed by a number of authors.  As indicated in Table~1, one of the key points is that the maximum confinement energy in GCs would be higher than that in galaxies because GCs have $\sim 0.1-1~\mu {\rm G}$ magnetic fields over $\sim 1$~Mpc~\footnote{Note that, when the coherent length is $\lambda_{\rm coh}$, CRs can be isotropized below $\sim e B \lambda_{\rm coh}^{1/2} r^{1/2}$, which is lower than $e B r$.  Thus, CRs with $E=eBr$ would be in the ballistic regime.  Also, the proton maximum energy achieved by accretion shocks is much smaller than the value in Table~1 since the shock velocity is much smaller than the light velocity and the magnetic field would be weaker at larger scales~\citep[e.g.,][]{uhecr,ino+05}.}.  

Recent gamma-ray observations have started to allow significant probes of dark matter~\cite{ack+10,hua+12,ale+10,abr+12,an12,arl+12} and CRs~\cite{ale+10,ack+10b,pin+11,jp11,arl+12} from GeV to multi-TeV energies, as expected from papers from the pre-\textit{Fermi} era~\cite{egret,pe04,an08}.  
In the GeV range, \textit{Fermi} has given interesting limits on models of dark matter and CRs.  In the TeV range, imaging atmospheric Cherenkov telescopes (IACTs) such as H.E.S.S., MAGIC, and VERITAS play an active part in detecting very-high-energy (VHE) gamma-ray sources~\cite{hh10}.  

Neutrino observations should also be relevant at high energies, especially at $\gtrsim 10$~TeV~\cite{ahr+04,icupdate,kat06}.  Neutrinos are special since they provide complementary probes of heavy dark matter and clues to confinement properties of high-energy CRs.  The IceCube detector at the South Pole was completed, and the comparable KM3Net detector in the Mediterranean sea is being planned.  Neutrinos have become one of the important and powerful messengers, along with gamma rays~\citep[e.g.,][]{mb12}, but there are only a few works about neutrinos from GCs in the contexts of both dark matter and CRs~\cite{ber+97,cb98,mur+08,yua+10} despite many on the gamma-ray emissions.  
Both gamma rays and neutrinos are essential for understanding high-energy processes in GCs.  This paper focuses on the power of current and future high-energy neutrino observations.   
 
Confirmations and tests of predicted signals are possible by multi-messenger detectors.  In particular, neutrinos provide a complementary probe even if gamma rays are more easily detected.  Neutrinos can probe hadronic signals.  For CRs, gamma rays may come from primary electrons, but neutrinos require protons or nuclei.  For dark matter, neutrinos infer channels involving decay into heavy leptons, quarks, massive bosons, Higgs, and neutrinos themselves.  Neutrino detectors have better sensitivities especially at higher energies.  At such high energies, gamma rays are attenuated or cascaded whereas neutrinos can leave a source without attenuation, which allows us to reveal injection energies directly.  Neutrino detectors typically have degree-scale angular resolution, and this is comparable to the size of nearby GCs.  While the large angular size means that the IACT sensitivity is worse than its point source sensitivity, neutrino detectors have some advantages in observing such extended sources.        
 
The organization of this paper is as follows. 
In Section~2, we overview gamma-ray and neutrino production processes in GCs, and describe basic concepts.  In Section~3, the results are shown.  Summary and discussion are found in Section~4.  We adopt $H_0 \equiv 100h =70.2~{\rm km}~{\rm s}^{-1}~{\rm Mpc}^{-1}$, $\Omega_{\rm dm}=0.229$, $\Omega_m=0.275$, and $\Omega_{\Lambda}=0.725$.

\section{Galaxy clusters as high-energy particle sources}
\subsection{Annihilating dark matter}
Many dark matter models have principal annihilation channels leading to quarks, W/Z gauge bosons, and Higgs.  One may consider leptonic or semi-leptonic decays.  For hadronic decays, many mesons are produced due to hadronization, leading to neutrinos, gamma rays and electrons/positrons via decay of mesons, and the gamma-ray spectrum from final states is near-universal though details depend on particle physics models~\citep[e.g.,][]{dm,pin+11,an12}.  
In addition, one may have internal bremsstrahlung, i.e., direct emission from a virtual, charged, exchanged particle~\citep[e.g.,][and references therein]{pin+11,bremss,ks07}.  The internal bremsstrahlung is particularly relevant if lowest-order annihilation processes are suppressed e.g., via a helicity or loop suppression.  It also allows electromagnetic emissions even for the $\nu \bar{\nu}$ channel in the tree level~\cite{bremss,ks07}.  

In this work, for demonstration purposes, we use a representative annihilation channel, $\chi \chi \rightarrow b \bar{b}$, where we run PYTHIA to evaluate the spectrum, $dS/dE$~\cite{sjo+06}.  Examples of electromagnetic and neutrino spectra are shown in Figures~1 and 2.  We also consider another representative annihilation channel, $\chi \chi \rightarrow \mu^+ \mu^-$, motivated by leptophilic models.  In those models, dark matter exclusively annihilates into a lepton and anti-lepton, which was extensively studied in the context of the PAMELA and \textit{Fermi} excess~\citep[see reviews][and references therein]{dm2}.  Such channels may be realized in some models such as a Sommerfeld-enhanced model including light force-scale carrier~\cite{ark+09}.  For demonstrative purposes, we simply use the well-known formula of three-body decay of muons.  The spectra are harder than those of channels involving hadronization.  

For comparison, spectra from other channels ($\chi \chi \rightarrow W^+ W^-$ and $\chi \chi \rightarrow t \bar{t}$) are also shown.  For the $W^+ W^-$ channel, the weak boson decays into a neutrino and lepton and can also be hadronic, while the top quark mainly decays into a weak boson and bottom quark for the $t \bar{t}$ channel.  In both the cases, the spectra are harder than that of the $\chi \chi \rightarrow b \bar{b}$ channel but softer than that of the $\mu^+ \mu^-$ channel, so the resulting constraints would lie in between the $b \bar{b}$ case and the $\mu^+ \mu^-$ case.  Also, we do not consider the internal bremsstrahlung since its relevance is model-dependent.  
  
Another particle physics factor is the self-annihilation cross section, $<\sigma v>_{\chi \chi}$, which we attempt to constrain in this work.  If dark matter is a thermal relic, the required cross section is $\approx (2-5) \times {10}^{-26}~{\rm cm}^3~{\rm s}^{-1}$, depending on mass~\cite{ste+12}.  However, larger or smaller cross sections are allowed if dark matter is nonthermally produced in the early universe.  
The cross section is unknown, but neutrino and gamma-ray limits give useful constraints on $<\sigma v>_{\chi \chi}$.    

\begin{figure*}[bt]
\begin{minipage}{0.49\linewidth}
\begin{center}
\includegraphics[width=\linewidth]{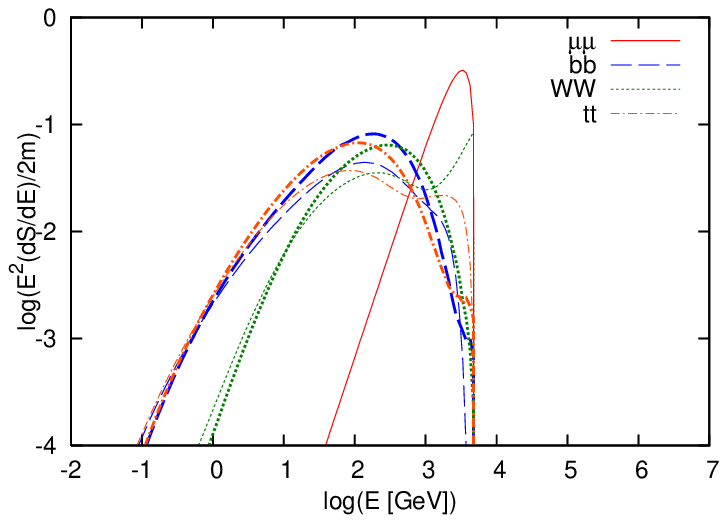}
\caption{
Injection spectra of gamma rays (thick curves) and electron/positrons (thin curves) from dark matter annihilation ($m_\chi c^2=5$~TeV) for $\chi \chi \rightarrow \mu^+ \mu^-$, $\chi \chi \rightarrow W^+ W^-$, $\chi \chi \rightarrow t \bar{t}$ and $\chi \chi \rightarrow b \bar{b}$.    
}
\end{center}
\end{minipage}
\begin{minipage}{.05\linewidth}
\end{minipage}
\begin{minipage}{0.49\linewidth}
\begin{center}
\includegraphics[width=\linewidth]{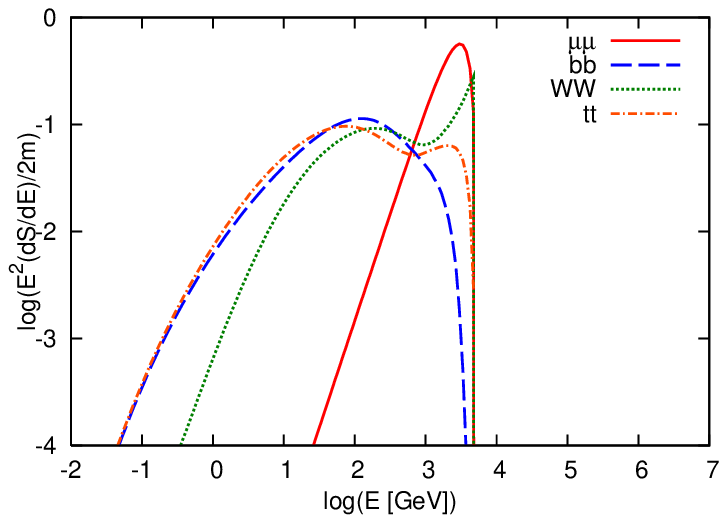}
\caption{
Injection spectra of neutrinos (sum of all flavors) from dark matter annihilation ($m_\chi c^2=5$~TeV) for $\chi \chi \rightarrow \mu^+ \mu^-$, $\chi \chi \rightarrow W^+ W^-$, $\chi \chi \rightarrow t \bar{t}$ and $\chi \chi \rightarrow b \bar{b}$. 
}
\end{center}
\end{minipage}
\end{figure*}

In order to calculate indirect signals of annihilation dark matter, we need to evaluate the astrophysical factor as well as the particle physics factor.  
The astrophysical factor is determined by the dark matter distribution, which is not certain at present.  The smooth dark matter profile is often described as~\cite{voi05,dm,yuk+07}
\begin{equation}
\rho_{\chi} (r) = \rho_s \frac{1}{{(r/r_s)}^{\alpha_2} {\left[ 1+ {(r/r_s)}^{\alpha_1} \right]}^{(\alpha_3-\alpha_2)/\alpha_1}}, 
\end{equation}
where $\rho_s$ and $r_s=r_{\Delta_c}/c_s$ are the scale density and scale radius, respectively.  Here $r_{\Delta_c}$ is the virial radius such that the enclosed dark matter mass is $M_{\Delta_c} = (4 \pi/3) (\Delta_c \rho_c) r_{\Delta_c}^3$, and $c_s$ is the mass-concentration parameter for given cluster mass.  Note that $\Delta_c$ is the overdensity relative to the critical density $\rho_c$~\cite{voi05}, and we use $\Delta_c=200$ throughout this paper.  Based on recent cosmological simulations, we also adopt the following mass-concentration parameter~\citep[e.g.,][]{duf+08},
\begin{equation}
c_s (M_{200},z) = \frac{7.85}{{(1+z)}^{0.71}} {\left( \frac{M_{200}}{2 \times {10}^{12} h^{-1} M_{\odot}} \right)}^{-0.081}.
\end{equation}
In Eq.~(2.1), $\alpha_1$ is the parameter that determines the shape at $r_s$, $\alpha_2$ is the inner slope, and $\alpha_3$ is the outer slope.  The Navarro-Frenk-White (NFW) profile is given by $\alpha_1=1$, $\alpha_2=1$ and $\alpha_3=3$~\cite{nfw95}, but different values are taken for other profiles such as Kravtsov~\cite{kra+98} and Moore~\cite{moo+99} profiles.  Recent simulations seem more consistent with shallower profiles such as the Einasto profile~\cite{ein65}, which also seems to be supported by some observations.  On the other hand, it is natural to expect that the dark matter profile is steepened by baryon dissipation so that annihilation signals are enhanced~\cite{an12,baryon}.  For reference, we consider the NFW profile throughout this work, but one should keep in mind that there is uncertainty in the smooth profile.    

Once the particle physics factor and astrophysical factor are evaluated, the intensity of gamma rays and neutrinos coming from the smooth dark matter halo (in the unit of particles per energy, area, time, and solid angle) can be written as 
\begin{equation}
I_E^{\rm sm} (\Theta) = \frac{{<\sigma v>}_{\chi \chi}}{8 \pi {(1+z)}^{2}} \frac{d S}{d E'} \int_{-l_{\rm max}}^{l_{\rm max}} dl \,\,\, \rho_{\chi}^2 (r), 
\end{equation}
where $r=\sqrt{l^2 + {(d_A \Theta)}^2}$, $l_{\rm max}=\sqrt{r_{\Delta_c}^2 - {(d_A \Theta)}^2}$, and $\Theta$ is the angular size measured from the direction to the GC center.  The normalization of the spectrum is given by $\Sigma \int d E \, E (dS/dE)=2 m_\chi c^2$.  
By performing integration over $\Theta$, the total flux is obtained as 
\begin{equation}
F_E^{\rm sm} = \frac{1}{4 \pi d_p^2} \frac{{<\sigma v>}_{\chi \chi}}{2 m_{\chi}^2}  \Delta_c \rho_{c} M_{\Delta_c} g_{M} (z) \frac{d S}{d E'},
\end{equation}
where we have introduced the flux multiplier for given cluster mass~\cite{ts03}, 
\begin{equation}
g_M (z) = \frac{V \int d V \, \rho_{\chi}^2 (r)}{{(\int dV \, \rho_{\chi} (r))}^2}.
\end{equation}
Here, $V$ is the volume of the GC.  The uncertainty in the smooth profile is included in this flux multiplier.  

In addition to the smooth dark matter distribution, there are contributions from dark matter substructures consisting of dark matter clumps.  To see interesting signals of annihilating heavy dark matter from GCs, we typically need significant boosts due to substructures, so that the smooth halo component is overwhelmed.  The expected boost is especially relevant for larger halos, including GCs.  The substructure model presented in Ref.~\cite{gao+11} gives
\begin{equation}
I_E^{\rm sub} (\Theta) = b(M_{200}) F_E^{\rm sm} \frac{1}{\pi \ln (17)} \frac{1}{\Theta^2 + {(\Theta_{200}/4)}^2}
\end{equation}
where $\Theta_{200}$ is the angular size corresponding to $r_{200}$.  We use the following boost factor,  
\begin{equation}
b(M_{200}) = 1.6 \times {10}^{-3} {\left( \frac{M_{200}}{M_{\odot}} \right)}^{0.39},
\end{equation}
where the dark matter clump size is assumed to be $M_{\rm min} = {10}^{-6} M_{\odot}$.  The boost factor also depends on the dark matter clump scale as $b \propto M_{\rm min}^{-0.2}$~\cite{gao+11}. 
Note that the total flux from dark matter annihilation is written as $F_E =(1+b) F_E^{\rm sm}$. 

The gamma-ray and neutrino angular profiles are relevant for the detectability.  To demonstrate this, we define the following quantity, 
\begin{equation}
J (\leq \Theta)= 2 \pi \int_0^{\Theta} d \Theta' \,\,\, I_E (\Theta'). 
\end{equation}
The results are shown in Figure~3, where substructures are included.  Note that $J(\leq \Theta)/F_E$ represents the normalized flux of the signal.  It is further divided by $\Theta$ since the background flux is proportional to $\Omega = 2 \pi (1- \cos \Theta) \approx \pi \Theta^2$, and we consider the signal relative to background uncertainties, i.e., the square root of the near-isotropic background intensity.  Figure~3 indicates how searches for the extended emission are crucial in the case of GCs.  The contribution from the smooth halo is concentrated in the central region, while that from the subhalos is much more extended.  For nearby clusters such as Virgo, in the presence of substructures, the apparent size is $\sim 1$~degree, which is comparable to the angular resolution of IceCube.  The optimization of the angular scale improves the detectability, though the best angular size depends on details of substructure models.  For example, in the case described in Eq.~(2.6), the signal from Virgo is maximized at $\sim 2-3$ degrees.

\begin{figure}[tb]
\begin{center}
\includegraphics[width=0.7\textwidth]{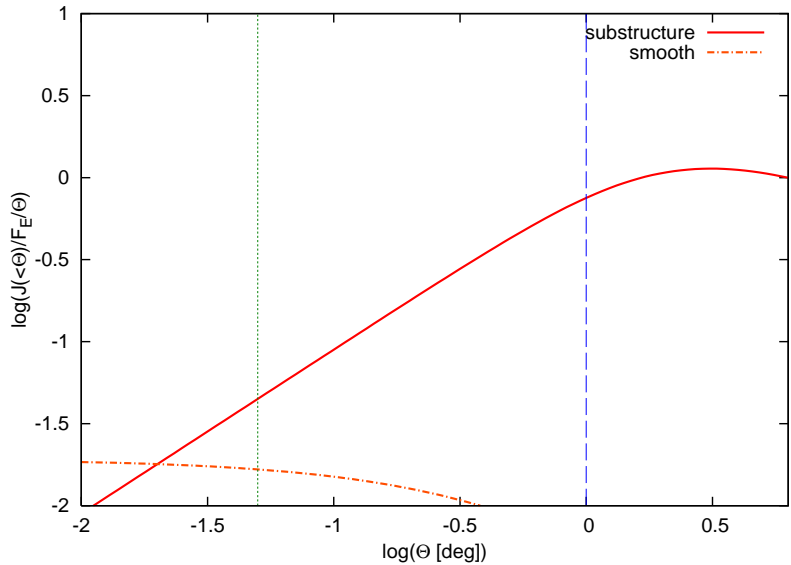}
\caption{The flux (within $\Theta$) divided by $\Theta$, $J (\leq \Theta)$, as a function of $\Theta$, which represents the shape of the signal to the square root of the background at given energy.  We assume $M_{\rm min}={10}^{-6} M_{\odot}$, where the contribution from substructures (thick curve) is largely dominant over the contribution from the smooth halo (dot-dashed curve).  For the dark matter profile, the parameter set for Virgo is used~\cite{san+11}. 
The typical angular resolution of IceCube ($\Theta=1$~deg) and CTA ($\Theta=0.05$~deg) are indicated by the dashed vertical line and the dotted vertical line, respectively.  The detectability is optimized at a few degrees in the presence of substructures.
}
\end{center}
\label{fig:f1}
\end{figure}

\subsection{Cosmic rays}
For astrophysical emissions, we focus on gamma-ray and neutrino production from the $pp$ reaction, i.e., interactions between CRs and target ICM gas.  This is enough as long as the CR spectral index is steep enough or the CR maximum energy is sufficiently less than $\sim {10}^{19}$~eV~\cite{mur+08,kot+09}.  The target gas density is related to the electron density, which is given by X-ray observations.  The $\beta$ model gives~\cite{voi05}
\begin{equation}
n_{e} (r) = n_{\rm co} {\left[ 1 + {\left(\frac{r}{r_{\rm co}}\right)}^{2} \right]}^{-3 \beta/2},
\end{equation}
where $n_{\rm co}$ is the core ICM density, $r_{\rm co}$ is the core radius and $\beta$ describes the slope.  The target nucleon density is determined by $n_N = n_e/(1-0.5x_{\rm He})$, where $x_{\rm He}=0.24$ is the Helium fraction~\citep[e.g.,][]{cb98,pe04}. 

The total CR energy, ${\mathcal E}_{\rm cr}$, is unknown, so it is introduced as a free parameter that we attempt to constrain.  The CR profile and spectrum depend on details of uncertain CR production mechanisms (including effects of the cluster assembly history), cooling processes, and transport processes via advection and diffusion~\citep[see, e.g.,][and references therein]{ens+07}.  In the case of nonrelativistic diffusive shock acceleration, the injection spectrum of CRs is expressed as $s_{\rm inj}=(\sigma+2)/(\sigma-1)$, where $\sigma$ is the compression ratio.  
A hard CR spectrum is achieved by high Mach number shocks that occur either at high redshifts during the formation of the protoclusters or today at the boundary where matter collapses from voids onto structured regions.  AGN and supernova remnants may also inject CRs with rather hard spectra, which may be important~\cite{fo12}, as well as the outflow from AGN~\cite{sut+11}.  Merger shocks with typical Mach numbers of ${\mathcal M} \sim 2-5$ lead to steeper spectra.  As noted above, the resulting CR spectrum inside GCs can be modified from the injection spectrum by various processes~\cite{pp10,kes10}.  

In this work, for demonstration purposes, we simply use a power law, $d S_{\rm cr}/d E_p \propto E_p^{-s}$, where $s$ is the CR spectral index.  One should keep in mind that the CR spectrum can be made concave by detailed physical processes such as the nonlinear CR acceleration~\citep[e.g.,][]{byk+08,vla+08}; but we avoid such model-dependent investigations. 
Then, the CR density $n_{\rm cr} (r)$ is connected to the total CR energy as 
\begin{equation}
{\mathcal E}_{\rm cr} = 4 \pi \int dr \,\,\, r^2 n_{\rm cr} (r) \int d {E'}_p \,\,\, {E'}_p \frac{d S_{\rm cr}}{d {E'}_p}. 
\end{equation}

The spatial distribution of CRs is quite uncertain at present, so we consider two extreme limits with the CR spectrum independent of $r$.  
In one limit, expecting that CR transport is dominated by advection, we assume that CRs trace the thermal energy distribution.  This would be the case especially for low-energy CRs.  One numerical simulation showed that the cluster assembly history leads to a CR spectral index of $s \sim 2.3$~\cite{pp10}.  For simplicity, we consider the isobaric model, where the CR energy density is a fraction of the thermal energy density, $X_{\rm cr} \equiv U_{\rm cr}/U_{\rm th}$.  Here $U_{\rm cr}$ is the energy density of CRs above 1~GeV, whereas $U_{\rm th} =(3/2)[1+(1-0.75 x_{\rm He})/(1-0.5 x_{\rm He})] n_e (r) k T_e (r)$ is the thermal energy density~\cite{pe04,an08}.   
In another limit, active CR transport is relevant, and we assume that CRs are homogeneously distributed over the virial radius~\footnote{See Model B in Ref.~\cite{mur+08}.  Ref.~\cite{mur+08} also considered the one-zone model by estimating the ICM density at the shock radius (Model A), but the external shock radius may be rather large, $\sim 2 r_{\rm vir}$~\cite{ryusim,vaz+08}.}.  Such a situation may be realized at sufficiently high energies, especially in the $\gtrsim$~TeV range.  For example, Ref.~\cite{kes10} suggested such a homogeneous distribution with $s \sim 2.2$ based on the radio and X-ray connection.  In particular, in the scenario where CRs are accelerated up to VHE or UHE by accretion shocks~\cite{uhecr,ino+05,mur+08}, a CR population with a hard spectral index is expected, though weak shocks would dominate the gravitational energy dissipation in GCs.  Note that the isobaric model can be regarded as an optimistic case in the sense that more gamma rays and neutrinos are produced compared to the uniform model.  

CRs are injected over a finite time, and spatial diffusion in GCs may lead to a break in the CR spectrum.  
When acceleration widely happens in GCs (which is especially the case for CRs produced by structure formation shocks), the spectral softening due to the diffusion is relevant above the critical energy where the diffusion time is comparable to the injection time~\citep[see, e.g.,][]{mur+08}.  Even below the critical energy, when the cluster size is much larger than that of CR accelerators, the diffusion can make the steady-state spectrum steeper by $1/3$ for the Kolmogorov turbulence, i.e., $s=s_{\rm inj}+1/3$, though the spatially-integrated spectrum may recover $s_{\rm inj}$.  In particular, if CRs diffuse without significant cooling after injections by a central point source such as AGN in the GC core, one obtains $n_{\rm cr} \propto r^{-1}$ if the steady state is achieved~\cite{ber+97,kot+09}.  In this case, depending on the CR energy, the upper limit of integration in Eq.~(2.10) should be $r_{\rm max}={\rm min}[r_{\rm diff},r_{\Delta_c}]$, where $r_{\rm diff} \equiv v_{\rm diff} t_{\rm inj}$ is the length for CRs to diffuse in the CR injection time $t_{\rm inj}$ and $v_{\rm diff}$ is the diffusion velocity.  We expect that the realistic situation would be between these two limiting cases.

One purpose of this work is to demonstrate how we can constrain this parameter, i.e., the total CR energy, in a less model-dependent way through high-energy neutrino observations.  The total CR energy can be related to the CR injection luminosity.  For accretion/merger shocks, the accretion luminosity is $L_{\rm ac} \sim {10}^{45}-{10}^{46}~{\rm erg}~{\rm s}^{-1}$~\cite{lw00,ino+05}, so the CR injection luminosity can be $L_{\rm cr} \sim {10}^{45}~{\rm erg}~{\rm s}^{-1}$ if $\sim 10$\% goes to CRs.  In addition, some AGN such as Fanaroff-Riley (FR) II galaxies, may have such high CR luminosities, and their luminosity could have been higher in the past.  Supernova remnants may have a comparable energy budget, although CRs may suffer from adiabatic losses.  In any case, the total CR energy stored in the GC is estimated to be 
\begin{eqnarray}
{\mathcal E}_{\rm cr} &\sim& L_{\rm cr} \times {\rm min}[t_{\rm inj},t_{\rm diff}] \nonumber \\
&\simeq& {10}^{61.5}~{\rm erg}~\left( \frac{L_{\rm cr}}{{10}^{45}~{\rm erg}~{\rm s}^{-1}}\right) \left( \frac{{\rm min}[t_{\rm inj},t_{\rm diff}]}{{\rm Gyr}}\right), 
\end{eqnarray}
where $t_{\rm inj}$ is the CR injection timescale, which is typically order of $\sim 1-10$~Gyr, and $t_{\rm diff}$ is the CR diffusion timescale. 

As in Eq.~(2.3), the intensity (per unit energy, area, time and solid angle) is given by 
\begin{equation}
I_E (\Theta)=  \frac{1}{4 \pi {(1+z)}^2} \int_{-l_{\rm max}}^{l_{\rm max}} dl \,\,\, n_N (r) n_{\rm cr} (r) \int d {E'}_p \,\,\, \frac{d S_{\rm cr}}{d {E'}_p}  \frac{d \sigma_{pp} \xi}{d E'} c
\end{equation}
where $\xi$ is the effective multiplicity of gamma rays or electrons/positrons or neutrinos, $c$ is the relative velocity between an incident proton and a target nucleon, and the CR normalization is given by $\int d E_p \, (dS_{\rm cr}/dE_p)=1$.  
Here, $l_{\rm max}=\sqrt{r_{\Delta_c}^2 - {(d_A \Theta)}^2}$ and $\Theta=r/d_A$ is the angular size measured from the cluster center.  
Examples of electromagnetic and neutrino spectra are shown in Figures~ 4 and 5.  The calculation is performed based on SYBILL~\cite{kel+06}.

\begin{figure*}[bt]
\begin{minipage}{0.49\linewidth}
\begin{center}
\includegraphics[width=\linewidth]{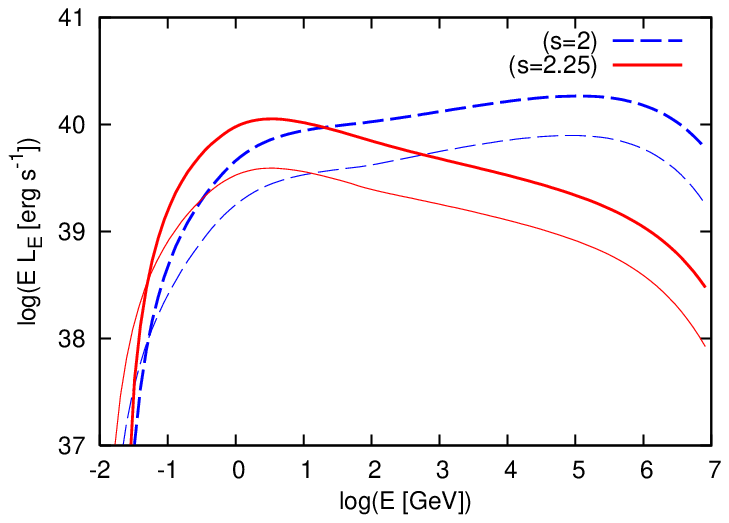}
\caption{Injection spectra of gamma rays (thick curves) and electron/positrons (thin curves) produced by CRs via the $pp$ reaction.  The total CR energy is assumed to be ${\mathcal E}_{\rm cr}={10}^{61}$~erg for CR spectral indices of $s=2$ and $s=2.25$, and a uniform CR distribution is used.}
\end{center}
\end{minipage}
\begin{minipage}{.02\linewidth}
\end{minipage}
\begin{minipage}{0.49\linewidth}
\begin{center}
\includegraphics[width=\linewidth]{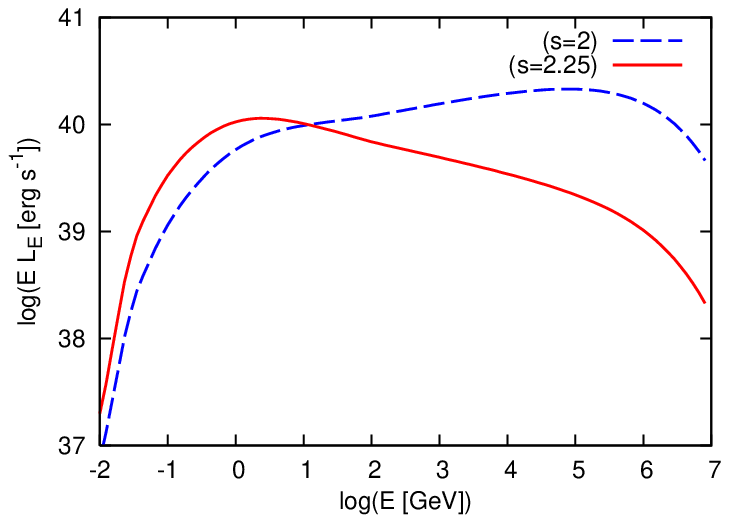}
\caption{Injection spectra of neutrinos (sum of all flavors) produced by CRs via the $pp$ reaction.  The total CR energy is assumed to be ${\mathcal E}_{\rm cr}={10}^{61}$~erg for CR spectral indices of $s=2$ and $s=2.25$, and a uniform CR distribution is used.
}
\end{center}
\end{minipage}
\end{figure*}

We evaluate the $J$ factor for the CR case.  As in the dark matter case, it depends on assumptions about the CR distribution.  For CRs, the $J$ factor defined by Eq.~(2.8) has the following dependences, 
\begin{eqnarray}
\frac{2 \pi \Theta I_E^{\rm uni}}{F_E^{\rm uni}} &\propto& \int dl \,\,\, n_e (r),\\
\frac{2 \pi \Theta I_E^{\rm iso}}{F_E^{\rm iso}} &\propto& \int dl \,\,\, k T_e (r) n_e^2 (r).
\end{eqnarray}
The results for the two CR profiles are shown in Figure~6.  As shown, the best angular scale is model-dependent.  For center-distributed CR profiles, the optimization should be done with smaller values of $\Theta$.  However, for the uniform CR distribution, the optimized angular scale is a few degrees.  

Though we focused on the $pp$ reaction in this work, if the CR spectral index is hard enough and that the proton maximum energy is sufficiently high, the Bethe-Heitler process is important~\cite{ino+05,kot+09}.  Such limiting cases will be studied in future work.

Also, one should keep in mind that gamma rays are produced via IC emission by electrons accelerated at accretion shocks~\cite{merger,lw00,kw09}.  Although the gamma-ray limit would give interesting constraints on the acceleration efficiency of electrons, it is beyond the scope of this work since our purpose is to demonstrate the power of neutrinos in constraining cosmic-ray ions in GCs. 

\begin{figure}[tb]
\begin{center}
\includegraphics[width=0.7\textwidth]{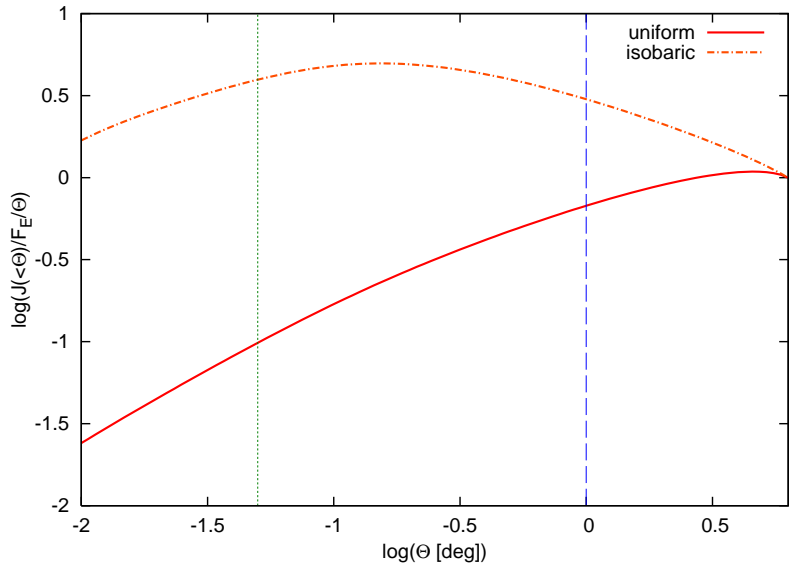}
\caption{The $J$ factor as a function of $\Theta$, which represents the shape of the signal to the square root of the background at given energy, but now for the CR case.  Two representative CR distributions are used, and the parameter set for Virgo is used.  
The typical angular resolution of IceCube ($\Theta=1$~deg) and CTA ($\Theta=0.05$~deg) are indicated by the dashed vertical line and the dotted vertical line, respectively.  The detectability is optimized at a few degrees for a uniform distribution.}
\end{center}
\label{fig:f1}
\end{figure}

\subsection{Neutrino and gamma-ray emissions}
The formulas for calculating the injection spectra of neutrinos and gamma rays are given by Eqs.~(2.3), (2.6) and (2.12).  Before presenting results from dedicated numerical calculations, it is useful to give rough estimates.  For annihilating dark matter, from Eqs.~(2.4) and (2.6), one has (for gamma rays or neutrinos)  
\begin{eqnarray}
E^2 F_E &\approx& \frac{1+b}{4 \pi d^2} \frac{{<\sigma v>}_{\chi \chi}}{m_{\chi} } (200 \rho_c) \frac{M_{200} c^2}{\mathcal R} g_M  \nonumber \\
&\simeq& 2.3 \times {10}^{-10}~{\rm GeV}~{\rm cm}^{-2}~{\rm s}^{-1}~\left( \frac{{<\sigma v>}_{\chi \chi}}{3 \times {10}^{-26}~{\rm cm}^{3}~{\rm s}^{-1}} \right) {\left( \frac{{m}_{\chi} c^2}{5~{\rm TeV}} \right)}^{-1} \nonumber \\ 
&\times& \left( \frac{M_{200}}{5 \times {10}^{14}~M_{\odot}} \right)  \left( \frac{(1+b) g_M}{{10}^4} \right) \left( \frac{10}{{\mathcal R}}\right) {\left( \frac{16~{\rm Mpc}}{d} \right)}^2, 
\end{eqnarray}
where ${\mathcal R} \equiv {\mathcal R} (E)= {(E^2 \frac{dS}{dE})}^{-1} \int dE \, (E \frac{dS}{dE})$ is the ratio of the bolometric flux to the differential flux at $E$ (see Figures~1 and 2).  

For CRs, Eq.~(2.12) leads to (for gamma rays or neutrinos)   
\begin{eqnarray}
E^2 F_E &\approx& \frac{1}{8 \pi d^2} f_{pp} \frac{L_{\rm cr}}{{\mathcal R}_{\rm cr}}
\approx \frac{1}{8 \pi d^2} \bar{n}_N \kappa_\pi \sigma_{pp} c \frac{{\mathcal E}_{\rm cr}}{{\mathcal R}_{\rm cr}} \nonumber \\
&\simeq& 1.5 \times {10}^{-10}~{\rm GeV}~{\rm cm}^{-2}~{\rm s}^{-1}~\left( \frac{\bar{n}_N}{{10}^{-4}~{\rm cm}^{-3}} \right) \left( \frac{{\mathcal E}_{\rm cr}}{{10}^{61}~{\rm erg}} \right) \left( \frac{20}{{\mathcal R}_{\rm cr}} \right) {\left( \frac{16~{\rm Mpc}}{d} \right)}^2 
\end{eqnarray}
where ${\mathcal R}_{\rm cr} \equiv {\mathcal R}_{\rm cr} (E_p)= {(E_p^2 \frac{dS_{\rm cr}}{dE_p})}^{-1} \int dE_p \, (E_p \frac{dS_{\rm cr}}{dE_p})$ and we assume that the gamma ray or neutrino roughly carries a half the energy of the mesons.  Since CRs are not depleted by the $pp$ reaction, the $pp$ meson production efficiency is estimated to be $f_{pp} (E_p) \approx \bar{n}_N \kappa_{\pi} \sigma_{pp} c t_{\rm int}$, where $\kappa_{\pi} \sigma_{pp} \approx {10}^{-26}~{\rm cm}^2$ is the effective cross section of meson production, and $t_{\rm int}$ is the interaction time of CRs. 
If CRs are trapped in GCs, we expect $t_{\rm int} \approx t_{\rm inj}$, and the injection timescale may be the dynamical timescale of accretion/merger shocks, or the cluster age if CRs are continuously supplied by AGN inside the GC.  
If the CR escape becomes relevant, we expect $t_{\rm int} \approx {\rm max} [t_{\rm diff}, r_{200}/c]$, taking into account that the transport velocity is limited by ${\rm min} [v_{\rm diff}, c]$.  In the latter case, when the diffusion coefficient increases with energy, gamma-ray and neutrino spectra become steeper than injection CR spectra.   

From Eqs.~(2.15) and (2.16), dark matter and CRs may thus lead to comparable energy fluxes, even though both cases are affected by astrophysical uncertainties.  Both predictions are not far from ranges of current and future experiments such as \textit{Fermi}, CTA, and IceCube/KM3Net.

Neutrinos reach Earth without attenuation.  On the other hand, the fate of gamma rays is significantly affected by electromagnetic processes inside and outside the GCs. 
GCs typically have magnetic fields $\sim 1-10~\mu$G around the cluster center, while $\sim 0.1-1~\mu$G in the cluster outskirts~\citep[see, e.g.,][and references therein]{mag,sr11}.
Since the Larmor radius of electrons/positrons is typically small enough, pairs produced in the GCs are trapped in the GCs, and lose their energies via the inverse Compton (IC) and synchrotron emissions~\footnote{At UHE, the Larmor radius of pairs is rather large, and the synchrotron cooling length becomes much shorter.  Then, the ``synchrotron pair halo/echo" emission can be expected~\cite{mur12}.}.  
Sufficiently high-energy gamma rays may also generate pairs via pair creation with low-energy photons, since the mean free path of gamma rays for the cosmic microwave background (CMB) is less than $\sim$~Mpc at energies from $\sim 100$~TeV to $\sim 3$~EeV.
As a result, IC emission is typically expected at $E_{\gamma}^{\rm br} \approx (4/3) \gamma_e^2 \varepsilon_{\rm CMB} \simeq 3.4~{\rm GeV}~{(\gamma_e/{10}^6)}^2$, where $\varepsilon_{\rm CMB}$ is the typical CMB energy and $\gamma_e$ is the electron/positron Lorentz factor.  Then, attenuated and cascaded gamma rays leave the GCs.  They may be further affected in intergalactic space.  What happens highly depends on uncertain intergalactic magnetic fields for whether the intergalactic cascade component contributes to the observed emission.  To be conservative, we only take into account attenuation by the extragalactic background light (EBL)~\citep[see][and references therein]{fin+10}.  

In this work, the spectrum of cascaded gamma rays inside the GC is numerically obtained by solving the following Boltzmann equations, 
\begin{eqnarray}
\frac{\pd N_\gamma}{\pd t} &=& -N_{\gamma} R_{\gamma \gamma} - \frac{N_{\gamma}}{t_{\rm esc}}
+ \frac{\pd N_{\gamma}^{\rm IC}}{\pd t} 
+ \frac{\pd N_{\gamma}^{\rm syn}}{\pd t} 
+ Q_{\gamma}^{\rm inj},\\
\frac{\pd N_e}{\pd t} &=& \frac{\pd N_e^{\gamma \gamma}}{\pd t} 
-N_{e} R_{\rm IC} + \frac{\pd N_e^{\rm IC}}{\pd t} 
- \frac{\pd}{\pd E} [P_{\rm syn} N_e] 
+ Q_e^{\rm inj}, 
\end{eqnarray}
where 
\begin{eqnarray}
R_{\gamma \gamma} &=& \int d \varepsilon \frac{dn }{d \varepsilon}  \int \frac{d \mu}{2} \,\, \tilde{c} \sigma_{\gamma \gamma} (\varepsilon, \mu), \nonumber \\
R_{\rm IC} &=& \int d \varepsilon \frac{dn }{d \varepsilon} \int \frac{d \mu}{2} \,\, \tilde{c}  \sigma_{\rm IC} (\varepsilon, \mu), \nonumber \\
\frac{\pd N_\gamma^{\rm IC}}{\pd t} &=& \int d E' N_{e} (E') \,\, \int d \varepsilon \frac{dn }{d \varepsilon} \,\, \int  \frac{d \mu}{2} \,\, \tilde{c}   \frac{d \sigma_{\rm IC}}{d E_\gamma} (\varepsilon, \mu, E'), \nonumber \\
\frac{\pd N_e^{\gamma \gamma}}{\pd t} &=& \int d E' N_{\gamma} (E') \,\, \int d \varepsilon \frac{dn }{d \varepsilon} \,\, \int \frac{d \mu}{2} \,\, \tilde{c}  \frac{d \sigma_{\gamma \gamma}}{d E_e} (\varepsilon, \mu, E'), \nonumber \\
\frac{\pd N_e^{\rm IC}}{\pd t} &=& \int d E' N_{e} (E') \,\, \int d \varepsilon \frac{dn }{d \varepsilon}  \,\, \int \frac{d \mu}{2} \,\, \tilde{c}  \frac{d \sigma_{\rm IC}}{d E_e} (\varepsilon, \mu, E'). 
\end{eqnarray}
Here $\tilde{c} =(1-\mu) c$, $t_{\rm esc}=r_{200}/c$ is the photon escape rate, $P_{\rm syn}$ is the synchrotron energy loss rate, $N_\gamma$ and $N_e$ are photon and electron/positron number densities per energy decade, and $Q_\gamma^{\rm inj}$ and $Q_e^{\rm inj}$ are photon and electron/positron injection rates.  
For simplicity, we consider a one-zone model, assuming an average magnetic field of $0.3~\mu$G.  The calculation is performed for the injection time of $t_{\rm inj}=1$~Gyr, and we obtain an essentially steady-state spectrum.  
For infrared and optical photons in GCs, we simply use the EBL with 10 times enhancement, and we adopt the low-IR model of Ref.~\cite{kne+04}.  The spectral energy distribution of galaxies has a peak around eV in $E^2 F_E$, and it is $\sim 10-100$ times larger than that of the EBL at $\sim 0.5-1$~Mpc from the GC center (relevant for dark matter clumps or confined CRs)~\cite[see, e.g.,][]{tm12}.  Although this treatment is not fully accurate, it is sufficient for our purpose to demonstrate the power of neutrinos.   

Our results for the spectra from dark matter annihilation are shown in Figure~7 for cascaded gamma rays and Figure~8 for neutrinos.  In the case of $\chi \chi \rightarrow b \bar{b}$, the gamma-ray spectrum is dominated by gamma rays coming from hadronization, though there is a lower-energy tail due to IC cascades.  For the $\chi \chi \rightarrow \mu^+ \mu^-$ channel, the IC emission gives the dominant contribution.  In particular, CMB photons are upscattered by pairs coming from muon decay.   

Our results for the spectra from CRs are shown in Figures~9 for cascaded gamma rays and Figure~10 for neutrinos.  The cascade component is relevant only when the CR spectrum is hard. For $s=2.25$, cascade gamma rays are negligible in the GeV-TeV range, and they are relevant only below $\sim 0.1$~GeV.  For $s=2$, pionic gamma rays are still the most important, but the IC emission by pairs give moderate contributions, and its enhancement to the total gamma-ray flux is a factor of $\sim 2-3$.   

\begin{figure*}[bt]
\begin{minipage}{0.49\linewidth}
\begin{center}
\includegraphics[width=\linewidth]{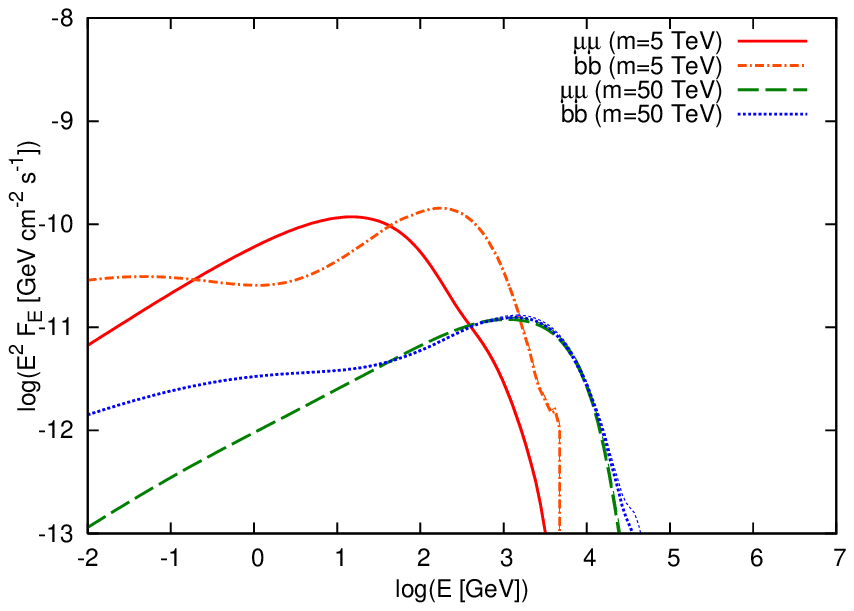}
\caption{
Spectra of cascaded gamma rays produced by annihilating dark matter.  The annihilation cross section is set to ${<\sigma v>}_{\chi \chi} = 2.2 \times {10}^{-26}~{\rm cm}^{3}~{\rm s}^{-1}$ with substructures.  The thick curves represent cases with EBL attenuation, while the thin curves are for cases without the EBL attenuation.  For the $b \bar{b}$ channel, there is a high-energy bump due to pionic gamma rays and a lower-energy tail due to IC cascades.  The Virgo cluster is considered, where the EBL attenuation is small.  The \textit{Fermi} limit is $E^2 F_E \sim 3 \times {10}^{-9}~{\rm GeV}~{\rm cm}^{-2}~{\rm s}^{-1}$~\cite{ack+10b}. 
}
\end{center}
\end{minipage}
\begin{minipage}{.05\linewidth}
\end{minipage}
\begin{minipage}{0.49\linewidth}
\begin{center}
\includegraphics[width=\linewidth]{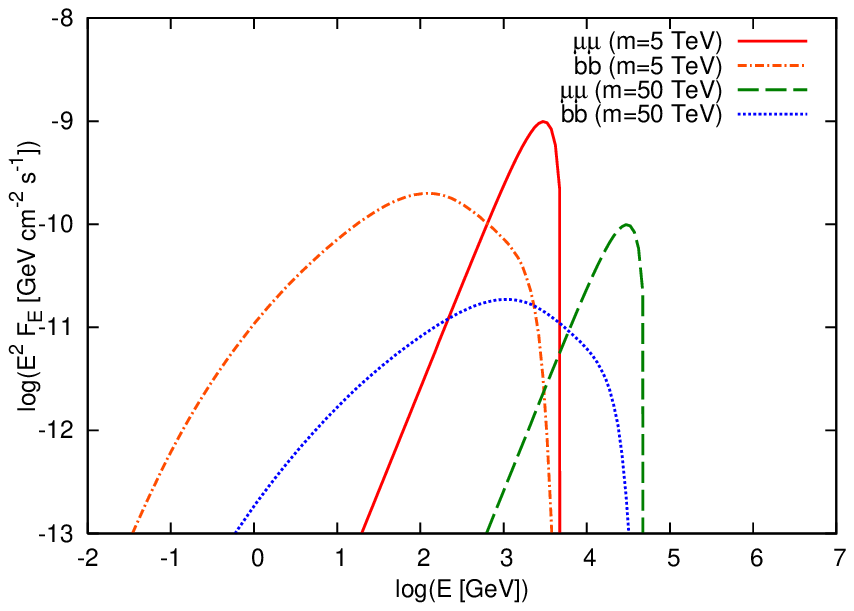}
\caption{
Spectra of neutrinos produced by annihilating dark matter.  The annihilation cross section is set to ${<\sigma v>}_{\chi \chi} = 2.2 \times {10}^{-26}~{\rm cm}^{3}~{\rm s}^{-1}$ with substructures.  The typical sensitivity of IceCube is $E^2 F_E \sim {10}^{-9}~{\rm GeV}~{\rm cm}^{-2}~{\rm s}^{-1}$. 
\newline \, \newline \, \newline \, \newline \, \newline \, \newline \, \newline}
\end{center}
\end{minipage}
\end{figure*}

\begin{figure*}[bt]
\begin{minipage}{0.49\linewidth}
\begin{center}
\includegraphics[width=\linewidth]{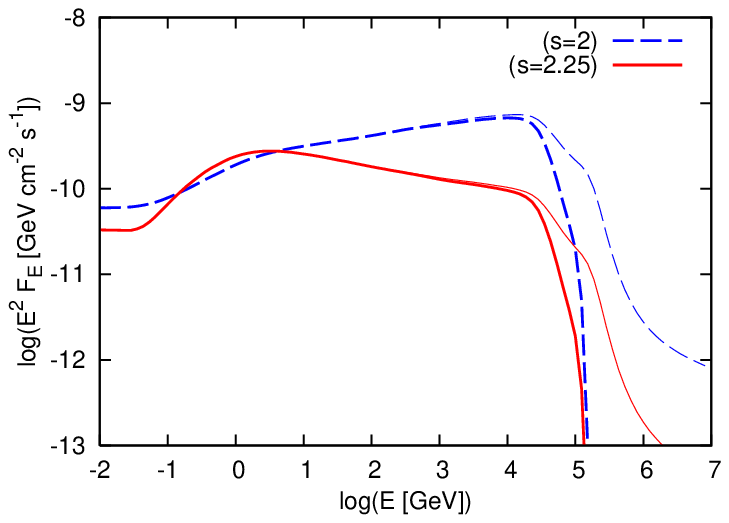}
\caption{
Spectra of cascaded gamma rays produced by CRs via the $pp$ reaction.  We use the same injection spectra shown in Figure~4.  The thick curves represent cases with EBL attenuation, while the thin curves are for cases without the EBL attenuation.  The \textit{Fermi} limit is $E^2 F_E \sim 3 \times {10}^{-9}~{\rm GeV}~{\rm cm}^{-2}~{\rm s}^{-1}$~\cite{ack+10b}.
}
\end{center}
\end{minipage}
\begin{minipage}{.05\linewidth}
\end{minipage}
\begin{minipage}{0.49\linewidth}
\begin{center}
\includegraphics[width=\linewidth]{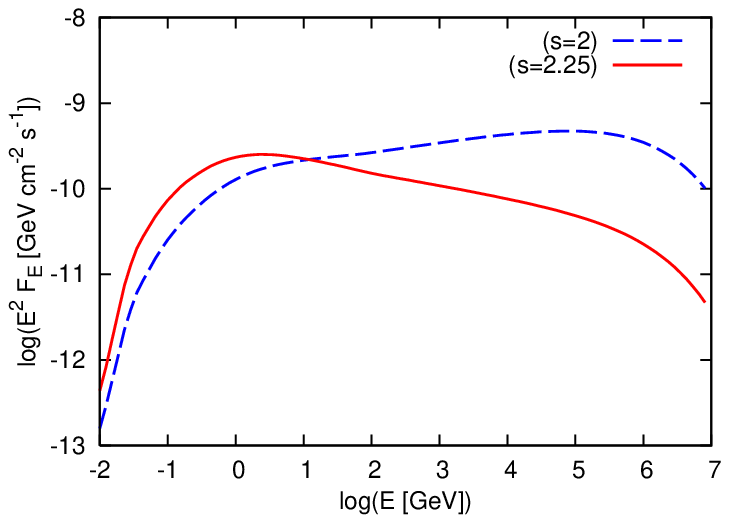}
\caption{
Spectra of neutrinos produced by CRs via the $pp$ reaction.  We use the same injection spectra shown in Figure~5.  The typical sensitivity of IceCube is $E^2 F_E \sim {10}^{-9}~{\rm GeV}~{\rm cm}^{-2}~{\rm s}^{-1}$.
\newline \, \newline \, \newline
}
\end{center}
\end{minipage}
\end{figure*}

\section{Neutrino and gamma-ray constraints}
In the previous section, we describe how we calculate neutrino and gamma-ray fluxes produced by annihilating dark matter and CRs.  Using these results, we derive neutrino constraints that could be placed by IceCube-like detectors in next several years, assuming the pessimistic case of non-detection.  To obtain the neutrino constraints, we estimate the muon event rate by
\begin{equation}
{\mathcal N}_{\mu} (\geq E_\nu) = \Omega T  \int dE_{\nu}^\prime \,\,\, A_{\rm eff} (E_{\nu}^\prime) \Phi_{\nu} (E_\nu^\prime), 
\end{equation}
where $\Omega$ is the observed solid angle, $T$ is the observation time, and $A_{\rm eff} (E_\nu)$ is the neutrino effective area averaged over the zenith angle (that is not fully accurate in the UHE range).  The full IceCube effective area is assumed to be three times as large as the IceCube-40 effective area~\cite{abb+11,abb+11b} with help of improvements in reconstruction techniques~\cite{icupdate}, and constraints are set by the criterion, ${\mathcal N}_{\rm sig}/\sqrt{{\mathcal N}_{\rm sig}+{\mathcal N}_{\rm bkg}} < 2$.  For the background ${\mathcal N}_{\rm bkg}$, we use the conventional atmospheric muon neutrino background~\cite{abb+11,bec08}.  For the signal ${\mathcal N}_{\rm sig}$, we consider neutrinos from annihilating dark matter or from CRs.  Neutrino mixing is taken into account assuming $\theta_{12}=0.59$ and $\theta_{23}=\pi/4$~\cite{sv06}.  At low energies (at least in the $\lesssim 300$~TeV range), the atmospheric neutrino flux is dominant, and the constraints will improve as the square root of time.  At higher energies, the atmospheric neutrino flux is negligible, and the constraints will improve linearly with time unless some other background is detected.  The above criterion would be conservative at higher energies.

We use $T=5$~yr and $\Omega = {\rm max}[\Theta^2, \pi \theta_{\nu}^2]$, where $\theta_\nu \approx 1.5 \sqrt{{\rm TeV}/E_\nu}$~deg is the kinematic angle.  We take $\Theta= 0.5 r_{\rm 200}/d_A$, which is appropriate for annihilating dark matter with substructures and CRs with a uniform distribution (see Figures~3 and 6).  Note that this gives conservative estimates for more center-distributed profiles. 

We only consider muon-track events.  For cascade events, depending on the criteria of event cuts, the effective area is typically smaller than that for muon-track events~\cite{abb+11c}.  Also, the angular resolution for cascade events is usually worse than that for muon-track events, which increases the atmospheric neutrino background though it can be lowered if muon neutrino charged-current events can be excluded~\citep[e.g.,][]{bc04,dl12}.  In principle, the situation can change if the angular resolution is improved in future neutrino detectors such as KM3Net (cf.~Ref.~\cite{dl12}).  Also, we do not consider any low-energy extension such as IceCube DeepCore~\cite{deepcore}, which can enhance the detectability below 100~GeV.  

Recently, gamma-ray constraints have been placed using \textit{Fermi} and IACT data.
Although we focus on neutrino constraints in this work, it is useful to make a comparison.  For that purpose, we show gamma-ray constraints taking into account electromagnetic cascades, from the flux upper limits used in Refs.~\cite{pin+11,ack+10b}.  Although we do not perform dedicated analyses, this is enough for our present purpose of showing the neutrino power, and calculating the electromagnetic cascades allows us to see how the gamma-ray constraints extend to very high masses. 

In this work, we select five nearby clusters, Virgo, Fornax, Perseus, Coma and Ophiuchus, using cluster parameters shown in Refs.~\cite{san+11,pe04}.  For cumulative neutrino backgrounds from all clusters, see Refs.~\cite{mur+08,mb12}.  We assume that no neutrino signal is observed in five years and consider either of the two contributions, annihilating dark matter or interacting CRs.  
In the future, some signal may be observed, and both may potentially contribute.  However, the predictions of each contribution are currently uncertain and model-dependent, so that it is sufficient to focus on setting limits on either of them without any signal.  If a signal is discovered, more detailed joint studies will be needed.

\subsection{Constraints on annihilating dark matter}
In Figure~11, we show forecasted neutrino constraints for five nearby clusters, assuming the $\mu^+ \mu^-$ channel.  The Virgo cluster will give the most stringent bound, where the constraint can be ${<\sigma v>}_{\chi \chi} \lesssim {10}^{-23.5}~{\rm cm}^3~{\rm s}^{-1}$ at $\sim 10$~TeV in five years.  For gamma-ray constraints, Virgo may not be a suitable target for the purpose of constraining dark matter because of contamination by gamma rays from the central AGN, M87.  However, neutrino signals have not been observed yet, so we do not have to take care about the astrophysical background until it is detected.  More distant clusters, Coma and Perseus, will give $\sim 10$ times weaker constraints, ${<\sigma v>}_{\chi \chi} \lesssim {10}^{-22.5}~{\rm cm}^3~{\rm s}^{-1}$ at $\sim 10$~TeV in five years.  For Ophiuchus and Fornax in the southern hemisphere, we assume a hypothetical detector in the northern hemisphere.  Our constraints agree with the independent work by Ref.~\cite{dl12} in the low-energy range, except that the neutrino effective area is slightly different and we consider the dependence of the atmospheric neutrino background on the zenith angle.  There is a difference by a factor of $\sim 3$ in the high-energy range, mainly because they took into account astrophysical neutrinos based on a model of Ref.~\cite{mur+08} as a background.    

In Figure~12, we show neutrino constraints for five clusters for the $b \bar{b}$ channel.  The constraints are much weaker than the results for the $\mu^+ \mu^-$ channel, because the spectrum due to hadronization is much softer than the spectrum due to lepton decay.  Similarly to the $\mu^+ \mu^-$ channel, the Virgo cluster gives the most stringent limits, where the constraint can be ${<\sigma v>}_{\chi \chi} \lesssim {10}^{-22}~{\rm cm}^3~{\rm s}^{-1}$ in five years.  Note that, in Figures~15 and 16, the conservative unitarity bound for $v_{\rm rms}=300~{\rm km}~{\rm s}^{-1}$ is also shown~\cite{bea+07}.  Note that the tighter bound of $m_\chi \lesssim 100$~TeV is obtained for thermal production in the early universe~\cite{annuni}.  The unitarity line displayed here is obtained in the late universe, when dark matter is non-relativistic.   Although a typical velocity for galaxy halos is chosen, this can be somewhat stronger for clusters (since the velocity dispersion of member galaxies is order of $\sim 800-1000~{\rm km}~{\rm s}^{-1}$).  Though even the conservative unitarity bound can be avoided if dark matter particles are not point-like, it typically gives the most stringent bound at sufficiently heavy masses, $\gtrsim 100$~TeV. 

Our constraints can be compared to other neutrino constraints. 
The IceCube-59 analyses for dwarf galaxies gave $\lesssim {10}^{-21}~{\rm cm}^{3}~{\rm s}^{-1}$ for the $\tau^+ \tau^-$ channel, which is weaker than the constraint for substructures but more model-independent~\cite{rot+11}.  The IceCube-22 analyses for the Galactic halo gave $\lesssim {10}^{-22}~{\rm cm}^{3}~{\rm s}^{-1}$ for the $\mu^+ \mu^-$ channel and $\lesssim {10}^{-20}~{\rm cm}^{3}~{\rm s}^{-1}$ for the $b \bar{b}$ channel, respectively~\cite{rot+11b}.  The IceCube-40 analyses for the Galactic center obtained $\sim 10$ times better constraints at $\sim 1$~TeV and comparable results at $\sim 10$~TeV~\cite{rot+11}.  In the presence of substructures with very small values of the dark matter clump mass, the extragalactic component can be more relevant in diffuse backgrounds.  For $M_{\rm min}={10}^{-6} M_\odot$, the corresponding average flux-multiplier is $g_0 \sim (1-3) \times {10}^6$, and one has $\lesssim {10}^{-23}~{\rm cm}^{3}~{\rm s}^{-1}$ for the $\mu^+ \mu^-$ channel and $\lesssim {10}^{-21.5}~{\rm cm}^{3}~{\rm s}^{-1}$ for the $b \bar{b}$ channel, respectively, at $\sim 30$~TeV~\cite{mb12}.   

We here point out an interesting implication of our results.  Recent preliminary \textit{Fermi} data have indicated the possible existence of the ``VHE Excess" in the $\sim 100$~GeV range~\cite{mur+12}.  This excess is attributed to cascaded gamma rays from astrophysical sources such as BL Lacs, but the dark matter interpretation could also be possible~\cite{mb12}.  However, for annihilating dark matter to explain the ``VHE Excess", quite large cross sections are required.  For $g_0 \sim 3 \times {10}^{6}$, one would roughly expect ${<\sigma v>}_{\chi \chi} \sim {10}^{-22.5}~{\rm cm}^3~{\rm s}^{-1}$ for the $\mu^+ \mu^-$ channel and ${<\sigma v>}_{\chi \chi} \sim {10}^{-23.5}~{\rm cm}^3~{\rm s}^{-1}$ for the $b \bar{b}$ channel, respectively.  However, the former case can be ruled out by neutrino observations of GCs, whereas the latter case is inconsistent with \textit{Fermi} gamma-ray constraints (see below).  However, accounting for the ``VHE Excess" by decaying dark matter is still allowed.      

\begin{figure*}[bt]
\begin{minipage}{0.49\linewidth}
\begin{center}
\includegraphics[width=\linewidth]{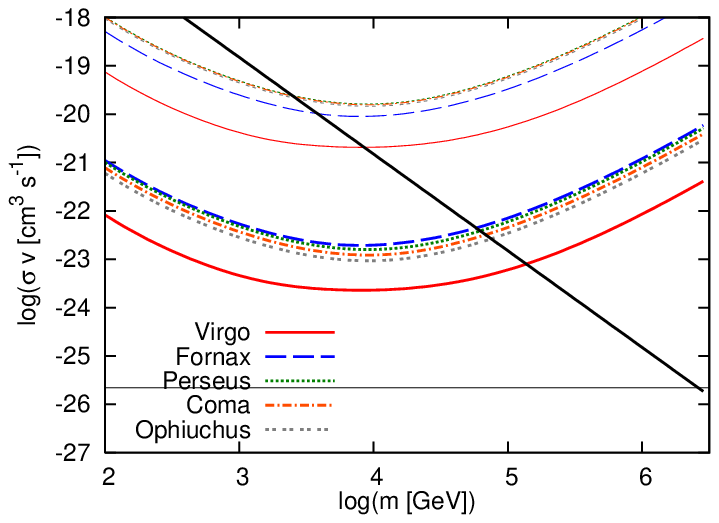}
\caption{Forecasted neutrino constraints on the dark matter annihilation cross section, ${<\sigma v>}_{\chi \chi}$, for five nearby GCs, with (thick curves) and without (thin curves) substructures.  Nondetection in five years is assumed for the $\mu^+ \mu^-$ channel.  The unitarity bound (thick solid curve) and the typical cross section of thermal relics (thin solid curve) are also shown.           
}
\end{center}
\end{minipage}
\begin{minipage}{.05\linewidth}
\end{minipage}
\begin{minipage}{0.49\linewidth}
\begin{center}
\includegraphics[width=\linewidth]{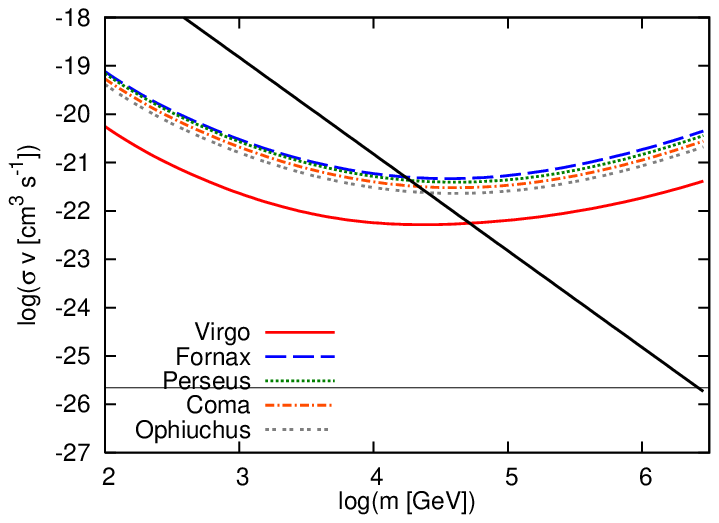}
\caption{The same as Figure~11, but for the $b \bar{b}$ channel.  Only the constraints with substructures are shown since the constraints without them are quite weak. 
\newline \, \newline \, \newline \, \newline 
}
\end{center}
\end{minipage}
\end{figure*}

In Figures~13 and 14, we show our results on two individual clusters, Fornax and Virgo, respectively.  
The neutrino constraints are the same as those shown in Figures~11 and 12.  We also show gamma-ray constraints (with cascades) obtained from \textit{Fermi} data, although our limits are relatively conservative.  Note that the latest gamma-ray constraints obtained by Refs.~\cite{hua+12,han+12} are better (though the constraints by Refs.~\cite{hua+12,han+12} is stronger than that by Ref.~\cite{an12} partially due to different choices of $\Delta_c$).    
Importantly, because the cascade effect is included in the calculation, our constraints extend to higher masses, even at $\gtrsim 10$~TeV, and one sees that the constraints become weaker $\propto m_{\chi}$.  

The gamma-ray constraints are more stringent than the neutrino bound in the typically-considered range of the WIMP mass, $\lesssim 100$~TeV.  In particular, for the $b \bar{b}$ channel, the neutrino bound is almost always weaker.  Even for the $\mu^+ \mu^-$ channel, the gamma-ray constraints are more stringent below $\sim 10$~TeV, but the neutrino constraints become stronger at higher masses.  

Neutrinos are often useful to set a reasonable limit on the total annihilation cross section when the branching ratios to various final states are unknown.  For some ``best" assumed final state, e.g., typically 100~\% gamma rays, the limit on the cross section will be stronger than for any other possible final state (with only standard model particles).  For some ``worst" assumed final state, e.g., typically 100~\% neutrinos, the limit on the cross section will be weaker than for any other possible final state.  If the branching ratios for all possible final states were known, the resulting cross section limit would be intermediate between these two extremes.  In other words, if the branching ratio to neutrinos were actually less than 100~\%, then the missing part would be going to final states that are easier to detect, and the true limit is stronger than assumed.  Hence, assuming that neutrinos are the ``worst" final state, the neutrino limit can be used as a reasonable limit on the total cross section (ignoring details at the level of a factor of 2, as typically one is trying to set limits on a log scale)~\cite{yuk+07,bea+07}.  When neutrinos are not the ``worst" final state, the logic can easily be modified~\cite{mb12}.  In our cases, the neutrino constraints for the $\mu^+ \mu^-$ channel are comparable to those for $\nu \bar{\nu}$ channel and are almost the weakest at $\lesssim 10$~TeV, where neutrino observations give a reasonable constraint on the total annihilation cross section.  

In Figure~14, for comparison, we also show the forecasted TeV gamma-ray constraint that can be placed by a future IACT, the Cherenkov Telescope Array (CTA)~\cite{act+11}.  This indicates that the potential TeV gamma-ray constraints are also powerful, and it will provide better constraints for the $b \bar{b}$ channel compared to the neutrino constraint. 
Here, for simplicity, the sensitivity is reduced by $\Theta/\Theta_{\rm ps}$ compared to the point source sensitivity, assuming the background-dominated case.  We do not provide more detailed studies for such extended sources since the detailed performance of IACTs needs to be known, but it will be done in the future when one better knows the effective area, point spread function and backgrounds.

\begin{figure*}[bt]
\begin{minipage}{0.49\linewidth}
\begin{center}
\includegraphics[width=\linewidth]{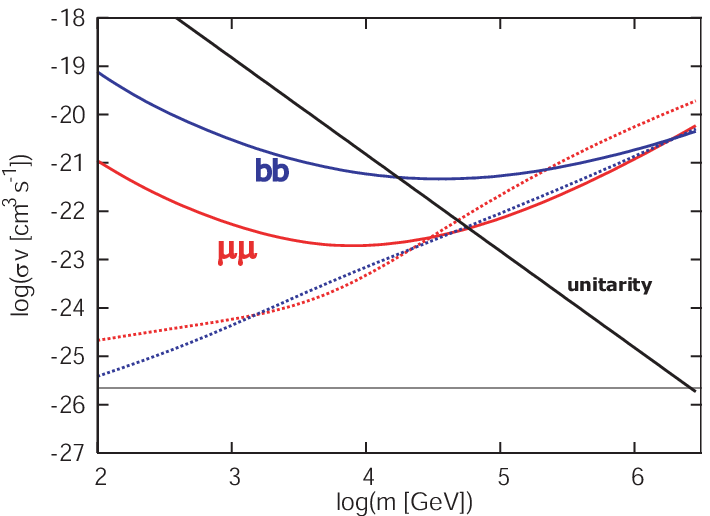}
\caption{Forecasted neutrino constraints (thick solid curves) and \textit{Fermi} gamma-ray constraints with cascades (thick dotted curves) on the dark matter annihilation cross section, ${<\sigma v>}_{\chi \chi}$, for the Fornax cluster.  The unitarity bound (thick solid curve) and the typical cross section of thermal relics (thin solid curve) are also shown.       
}
\end{center}
\end{minipage}
\begin{minipage}{.05\linewidth}
\end{minipage}
\begin{minipage}{0.49\linewidth}
\begin{center}
\includegraphics[width=\linewidth]{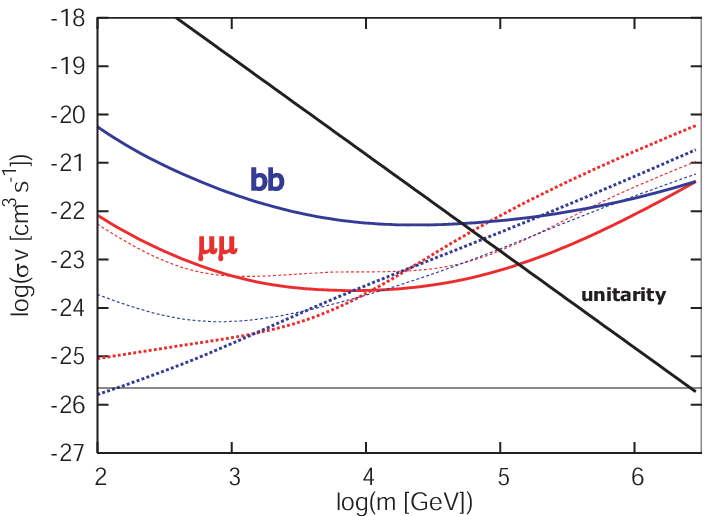}
\caption{The same as Figure~13, but for the Virgo cluster.  In addition, we overlay future gamma-ray constraints that can be placed by CTA (thin dotted curves).
\newline \, \newline \, \newline
}
\end{center}
\end{minipage}
\end{figure*}

Our forecasted neutrino constraints can be compared to other gamma-ray constraints. 
For example, \textit{Fermi} constraints from dwarf galaxies at $\sim 1$~TeV are $\lesssim {10}^{-22.5}~{\rm cm}^{3}~{\rm s}^{-1}$ for the $\mu^+ \mu^-$ channel and $\lesssim {10}^{-24}~{\rm cm}^{3}~{\rm s}^{-1}$ for the $b \bar{b}$ channel, respectively~\cite{dwarfann}.  The constraints become weaker $\propto m_{\chi}$.  The diffuse background measurements typically lead to weaker constraints~\cite{mb12,isodm,aba+10}.  Other limits from IACT observations also give interesting constraints~\citep[see a review][and references therein]{dm2}. 
For the NFW profile, in the TeV range, the recent HESS measurement gave $\lesssim {10}^{-21.5}~{\rm cm}^{3}~{\rm s}^{-1}$ for both the channels~\cite{abr+12}.  With significant boosts, $\lesssim {10}^{-23}~{\rm cm}^{3}~{\rm s}^{-1}$ for the $b \bar{b}$ channel was obtained.  For the galactic center, a stronger constraint of $\lesssim 3 \times {10}^{-25}~{\rm cm}^{3}~{\rm s}^{-1}$ is placed at TeV for the quark-antiquark channel~\cite{abr+11}.  Note that the isotropic DGB is difficult for IACTs to measure.  

Although we focused on the neutrino emission, the search for gamma-ray lines has been performed~\cite{annline}.  Recently, the existence of the line has been claimed around the Galactic center~\cite{130gev}, and GCs may also be interesting sources~\cite{hek+12}.

\subsection{Constraints on cosmic rays}
In Figure~15, we show the results of neutrino constraints on the total CR energy, ${\mathcal E}_{\rm cr}$, assuming a uniform CR distribution.  One sees that the neutrino bound is more stringent for harder CR spectral indices.  This is simply because IceCube/KM3Net is most sensitive to $\sim$~PeV neutrinos, where the atmospheric neutrino background is expected to be small.  For spectral indices, $s \gg 2$, most of the CR energy is concentrated in the GeV range, so that the CR energy in the PeV range is so small that the neutrino constraints should be weak.  
One sees that the Virgo cluster gives the most stringent neutrino constraint, ${\mathcal E}_{\rm cr} \lesssim {10}^{62}$~erg for $s=2$.  

Neutrinos with $\sim$~PeV energies are produced by protons with $\sim 30$~PeV~\cite{mur+08}.  Although it might be difficult to trap such high-energy CRs in GCs, it is useful to consider the isobaric model as an optimistic case.  In this case, CRs are more clustered around the GC center, so the neutrino flux is enhanced for the same total CR energy.  In Figure~16, we show forecasted neutrino constraints on the CR energy fraction in the isobaric model, $X_{\rm cr}$.  More massive GCs are expected to be larger energy reservoirs and the neutrino flux is proportional to $n_N^2$ rather than $n_N$, so the order among the five clusters changes from that in Figure~15.  One sees that the Perseus cluster gives the most stringent neutrino constraint, $X_{\rm cr} \lesssim 0.03$ for $s=2$.  

\begin{figure*}[bt]
\begin{minipage}{0.49\linewidth}
\begin{center}
\includegraphics[width=\linewidth]{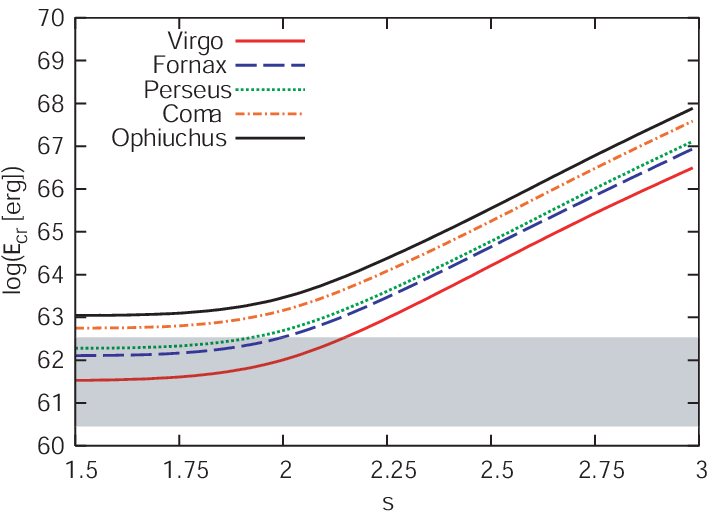}
\caption{
Forecasted neutrino constraints on the total CR energy, ${\mathcal E}_{\rm cr}$, for five nearby GCs.  The uniform CR distribution is assumed.  The Virgo cluster gives the most stringent constraint.  The shaded region indicates the typical total CR required in the scenario where GCs contribute to the observed CR flux.  
}
\end{center}
\end{minipage}
\begin{minipage}{.05\linewidth}
\end{minipage}
\begin{minipage}{0.49\linewidth}
\begin{center}
\includegraphics[width=\linewidth]{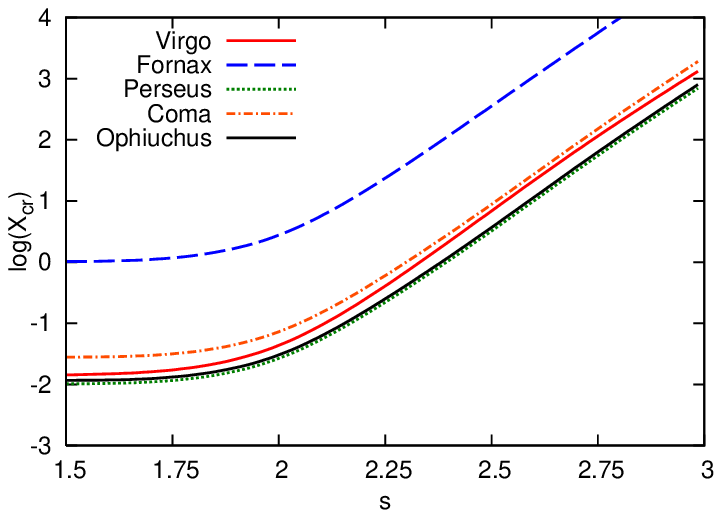}
\caption{
Forecasted neutrino constraints on the CR energy fraction in the isobaric model, $X_{\rm cr}$, for five nearby GCs.  The CR distribution is assumed to trace the thermal energy distribution.  One sees that the Perseus cluster gives the most stringent constraint. 
\newline
}
\end{center}
\end{minipage}
\end{figure*}

In Figures~17 and 18, we show our results of multi-messenger constraints on CRs for Perseus and Virgo, respectively~\footnote{We hereafter show the results up to $s=2.5$ for readability of the figures.  Neutrino constraints become too weak as the CR spectrum is softer than 2.5.}.  The neutrino constraints are the same as those shown in Figures~15 and 16.  To demonstrate the power of neutrinos, we also show gamma-ray constraints estimated from \textit{Fermi} data, taking into account cascades.  As expected, \textit{Fermi} gamma-ray limits are sensitive to GeV gamma rays, so the constraints from gamma rays are more stringent than the neutrino constraints for sufficiently steep CR spectral indices.  
For the Perseus cluster, the forecasted neutrino constraint is tighter than the gamma-ray constraint at $s \lesssim 2.3$.  The relative importance seems weaker for the Virgo cluster, but we stress that the neutrino constraint provides an independent constraint on CR ions since GeV-TeV gamma rays can be produced by primary electrons and the extent of CR clustering may be different between GeV and PeV-EeV.  Hence, independently of the gamma-ray limits, neutrino observations are important in order to constrain the amount of CRs that are accelerated by AGN or high-Mach number accretion shocks around the virial radius.  
In other words, our results suggest that the multi-messenger observations will also be able to give interesting constraints on specific models based on numerical simulations taking into account the cluster history, and details of CR acceleration and transport processes.  
 
In Figure~18, we show the forecasted TeV gamma-ray constraint that can be placed by CTA.  It is indicated that the TeV gamma-ray constraint is also powerful, and it will provide a better constraint for steeper CR spectral indices compared to the neutrino constraint. 
   
\begin{figure*}[bt]
\begin{minipage}{0.49\linewidth}
\begin{center}
\includegraphics[width=\linewidth]{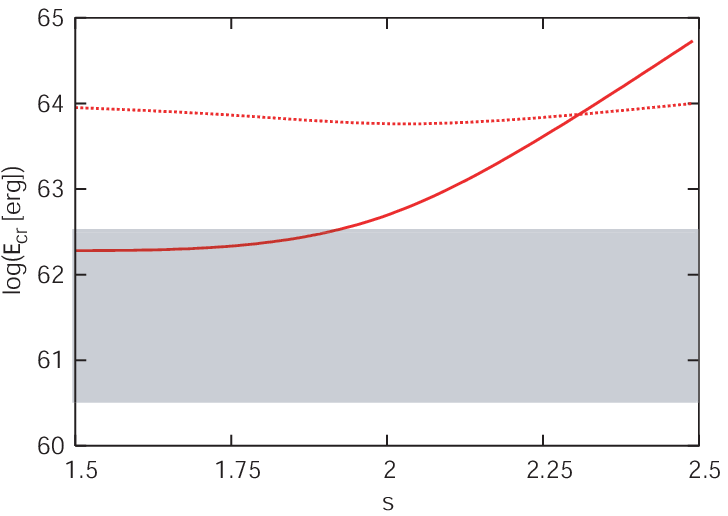}
\caption{The forecasted neutrino constraint (thick solid curve) and the \textit{Fermi} gamma-ray constraint with cascades (thick dotted curve) on the total CR energy, ${\mathcal E}_{\rm cr}$, for the Perseus cluster.  One sees that the neutrino constraint is stronger for harder indices.      
}
\end{center}
\end{minipage}
\begin{minipage}{.05\linewidth}
\end{minipage}
\begin{minipage}{0.49\linewidth}
\begin{center}
\includegraphics[width=\linewidth]{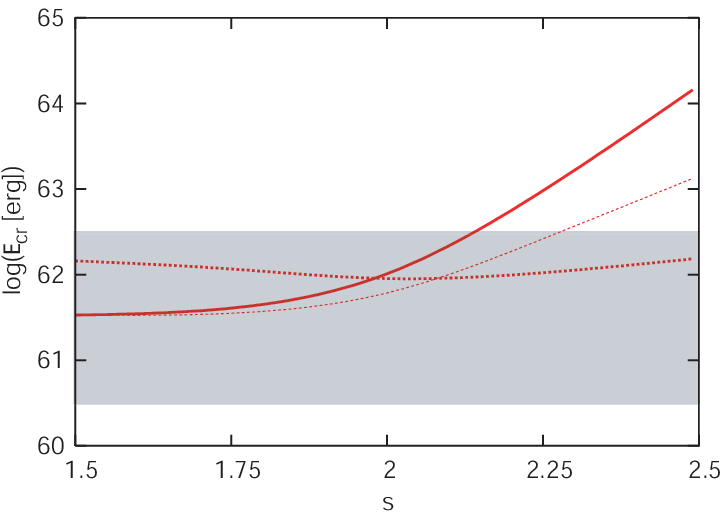}
\caption{The same as Figure~17, but for the Virgo cluster.  For comparison, we overlay future gamma-ray constraints that can be placed by CTA (thin dotted curve).
\newline
}
\end{center}
\end{minipage}
\end{figure*}
   
GCs could be sources of the observed CR flux.  For GC accretion shocks, assuming a total mass accretion rate is $\dot{M} \approx 0.1 V_{\rm ff}^3/G$ (where $V_{\rm ff}$ is the velocity of the infalling gas)~\cite{voi05}, the dissipation rate through the accretion shock is estimated to be $L_{\rm ac} \simeq 9 \times {10}^{45}~{\rm erg}~{\rm s}^{-1}~(f_g/0.16) M_{15}^{2/3}$, where $f_g = \Omega_b/\Omega_m$ is the gas fraction.  A fraction of the kinetic energy would dissipate by relatively high Mach-number shocks~\cite{ryusim,vaz+08}.  For AGN, the typical jet luminosities of FR I and II galaxies are $\sim {10}^{43}-{10}^{45}~{\rm erg}~{\rm s}^{-1}$ and $\sim {10}^{45}-{10}^{47}~{\rm erg}~{\rm s}^{-1}$, respectively~\cite{blazar}.  In either case, it is assumed that a significant fraction of the dissipated energy is used for CR acceleration.  On the other hand, the number density of GCs with masses above ${10}^{15} M_{\odot}$ is $n_{\rm gc} \approx 3 \times {10}^{-6}~{\rm Mpc}^{-3}$~\citep[e.g.,][]{jen+01}, but it becomes $n_{\rm gc} \approx$~a few~$\times {10}^{-5}~{\rm Mpc}^{-3}$ for masses above $5 \times {10}^{14} M_{\odot}$~\footnote{Hence the prediction for individual GCs given by Ref.~\cite{mur+08} is affected by the minimum mass of GCs that contribute to the observed CR flux, while the diffuse neutrino background prediction does not change much.}.  For GCs hosting AGN, only a fraction of GCs (and galaxy groups) would have powerful AGN, and $n_{\rm gc} \sim {10}^{-5}~{\rm Mpc}^{-3}$ is used in Ref.~\cite{kot+09}.  Then, taking into account the luminosity of CRs above ${10}^{17}$~eV is smaller than that above GeV by $\sim 5-1000$ (for $s \sim 2-2.4$), the energy budget of VHECRs may be 
\begin{equation}
L_{\rm vhecr} n_{\rm gc} \approx 3.2 \times {10}^{45}~{\rm erg}~{\rm Mpc}^{-3}~{\rm yr}^{-1}~\left( \frac{L_{\rm vhecr}}{{10}^{43}~{\rm erg}~{\rm s}^{-1}} \right) \left(\frac{n_s}{{10}^{-5}~{\rm Mpc}^{-3}} \right),
\end{equation} 
which can be comparable to the energy budget of observed CRs above $\sim {10}^{17}$~eV, $Q_{\rm vhecr} \approx 3 \times {10}^{45}~{\rm erg}~{\rm Mpc}^{-3}~{\rm yr}^{-1}$.  Then, the diffuse neutrino background flux can be order of $E_\nu^2 \Phi_\nu \sim {10}^{-9}-{10}^{-8}~{\rm GeV}~{\rm cm}^{-2}~{\rm s}^{-1}~{\rm sr}^{-1}$, which could be seen by IceCube/KM3Net~\cite{mur+08}. 

Next we briefly consider implications of future neutrino constraints on individual clusters.
For example, for $s=2.25$, the luminosity of injected CRs above above ${10}^{17}$~eV, $L_{\rm vhecr}={10}^{43}~{\rm erg}~{\rm s}^{-1}$, corresponds to $L_{\rm cr} \approx {10}^{45}~{\rm erg}~{\rm s}^{-1}$.  For $s=2$ below ${10}^{17}$~eV and $s=2.5$ above ${10}^{17}$~eV~\cite{mur+08}, the corresponding luminosity becomes $L_{\rm cr} \approx {10}^{44}~{\rm erg}~{\rm s}^{-1}$.   Then, through Eq.~(2.11), the total CR energy amount may be ${\mathcal E}_{\rm cr} \approx {10}^{60.5}-{10}^{62.5}$~erg (see shaded areas in Figure~15, 17-19).  Although details depend on the history of CR acceleration and escape properties, this implies that neutrino observations could test scenarios such that GCs contribute to the observed CR flux below the ankle.  Note that only optimistic cases would be probed by IceCube/KM3Net via the search for individual steady sources, but stacking analyses can improve the situation.  In addition, the diffuse background flux limit would give powerful and useful constraints~\cite{mur+08}.  

In Figure~19, we show the case of lower and higher values of the proton maximum energy. 
For lower maximum energies, the constraint becomes weaker, since the atmospheric neutrino background gets more important.  For higher maximum energies, the constraint does not change in the interesting range of the spectral index, $s \gtrsim 2$, since the neutrino flux at sufficiently high energies is almost the same.  Note that, when the maximum energy is high enough, the constraint for $s \lesssim 2$ is optimistic due to severe attenuation in Earth. 

Finally, in Figure~20, we show the results for the CR energy fraction in the isobaric model, $X_{\rm cr}$, for the Virgo cluster.  Again, independently of the gamma-ray limits, neutrino observations can provide constraints on the energy fraction of CRs that are accelerated by AGN or high-Mach number accretion shocks around the virial radius.      

\begin{figure*}[bt]
\begin{minipage}{0.49\linewidth}
\begin{center}
\includegraphics[width=\linewidth]{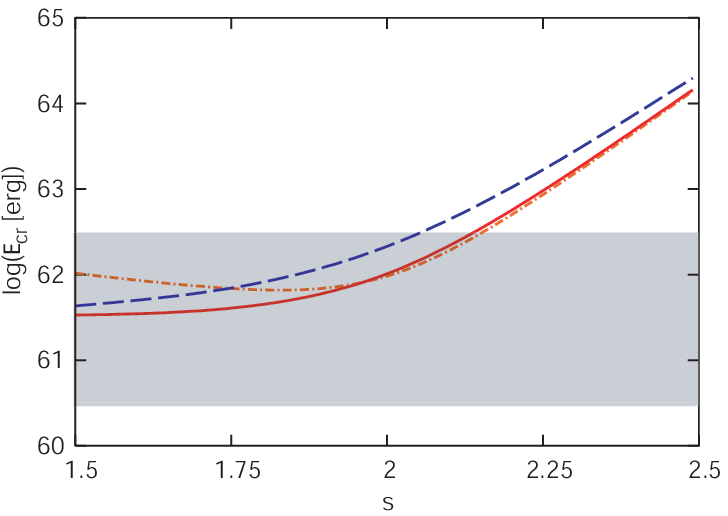}
\caption{The dependence on the proton maximum energy in neutrino constraints for the Virgo cluster.  The solid, dashed and dot-dashed curves are for ${10}^{17}$~eV, ${10}^{15.5}$~eV and ${10}^{19}$~eV, respectively.   
\newline
}
\end{center}
\end{minipage}
\begin{minipage}{.05\linewidth}
\end{minipage}
\begin{minipage}{0.49\linewidth}
\begin{center}
\includegraphics[width=\linewidth]{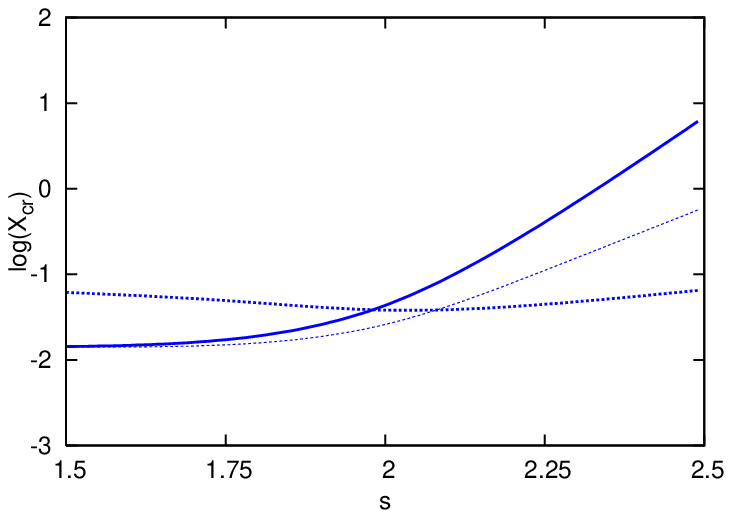}
\caption{The forecasted neutrino constraint (thick solid curve) and the \textit{Fermi} gamma-ray constraint with cascades (thick dotted curve) on the CR energy fraction in the isobaric model, $X_{\rm cr}$, for the Virgo cluster.  We also overlay future gamma-ray constraints that can be placed by CTA (thin dotted curve).  
}
\end{center}
\end{minipage}
\end{figure*}

\section{Summary and discussion}
In this work, we studied GCs as gigantic reservoirs of dark matter and CRs, and considered what we can learn from neutrino observations by IceCube and KM3Net, along with gamma-ray observations by \textit{Fermi} and IACTs.  Our work can be summarized as follows.   

(1) GCs could be interesting sources of neutrinos and gamma rays coming from annihilating dark matter, if the dark matter clump size is very small.  When the dark matter signal is boosted by substructures, the forecasted neutrino constraint for $\mu^+ \mu^-$ channel gives ${<\sigma v>}_{\chi \chi} \lesssim {10}^{-23}~{\rm cm}^3 {\rm s}^{-1}$ by five-year observations of IceCube or KM3Net, though it can be much weaker for other channels.  We also studied the effects of electromagnetic cascades on the gamma-ray constraints, which allows us to compare between the neutrino and gamma-ray constraints at very high masses.  For the $b \bar{b}$ channel, the neutrino constraint is almost always weaker than gamma-ray constraints.  For the $\mu^+ \mu^-$ channel, however, the neutrino constraint can be stronger than the gamma-ray constraints for $m_\chi c^2 \gtrsim 10$~TeV.  At lower masses, the gamma-ray constraints are stronger, where the neutrino constraint should be regarded as a reasonable bound on the total annihilation cross section.  

(2) GCs could be interesting sources of neutrinos and gamma rays since they are enormous reservoirs of CRs produced by accretion/merger shocks, hosted AGN and possibly supernova remnants.  High-energy neutrino observations will give constraints on the total CR energy of ${\mathcal E}_{\rm cr} \lesssim {10}^{62}$~erg with five-year observations of IceCube or KM3Net.  If VHECRs trace the ICM gas, $X_{\rm cr} \lesssim 0.1$ can be obtained for $s \lesssim 2.2$, independently of gamma-ray observations.  We also calculated the electromagnetic cascades, but they are typically irrelevant unless the CR spectrum is very hard.  Although the \textit{Fermi} gamma-ray constraint is stronger than the neutrino constraint for steep CR spectra, the neutrino limit is also powerful for CRs with hard spectra, which may be expected for CRs accelerated at accretion shocks or by AGN.  Our study demonstrated that not only gamma-ray observations but also neutrino observations give crucial constraints on the CR population in GCs, including the scenario where CRs from GCs contribute to the observed CR flux.  
Importantly, neutrino observations allow us to probe not electrons but CR ions, unlike gamma-ray observations.  Although we focus on the constraints assuming no signals are observed, possible detections in the near future may provide crucial clues to the nonthermal physics of GCs, the origin of CRs and so on.

Multi-messenger observations, neutrino observations combined with gamma-ray, X-ray and radio measurements~\citep[e.g.,][]{kes10,radio}, will definitely enable us to give deeper insights into heavy dark matter and CRs contained in GCs.  

When we were completing this draft, a related but independent study by Dasgupta and Laha was posted on arXiv:1206.1322. 

\section*{Acknowledgments}
K.M. is supported by JSPS and CCAPP.  The research of J.F.B. is supported by NSF Grant PHY-1101216.  We thank Basu Dasgupta, Shunsaku Horiuchi, Ranjan Laha, Paul Sutter, and especially Carsten Rott for helpful discussions.  We also thank an anonymous referee for valuable comments. 

\newpage

\bibliographystyle{JHEP}
\bibliography{ms}


\end{document}